\documentclass[authoryear,3p, preprint,review,12pt]{elsarticle}

\usepackage{amssymb}
\usepackage{amsmath}
\usepackage{multirow}
\usepackage{url}
\usepackage{graphicx}   % For \resizebox
\usepackage{float}      % For [H] placement
\usepackage{amssymb}    % For \checkmark
\usepackage{subcaption}
\usepackage{booktabs}
\usepackage{rotating} % For sidewaysfigure environment
\usepackage{siunitx}

\usepackage{graphicx}  % already present in most templates

\journal{Computers and Electronics in Agriculture}

\begin{document}

\begin{frontmatter}



\title{Integrating Feature Selection and Machine Learning for Nitrogen Assessment in Grapevine Leaves using In-Field Hyperspectral Imaging}


\author[inst1]{Atif Bilal Asad\corref{cor1}\fnref{fn1}}
\author[inst1]{Achyut Paudel}
\author[inst1]{Safal Kshetri}
\author[inst3]{Chenchen Kang}
\author[inst4]{Salik Ram Khanal}
\author[inst5]{Nataliya Shcherbatyuk}
\author[inst8]{Pierre Davadant}
\author[inst6]{R. Paul Schreiner\fnref{fn2}}
\author[inst7]{Santosh Kalauni}
\author[inst2]{Manoj Karkee}
\author[inst5]{Markus Keller}

\affiliation[inst1]{organization={Center for Precision and Automated Agricultural Systems},%Department and Organization
            addressline={Washington State University}, 
            city={Prosser},
            postcode={99350}, 
            state={WA},
            country={USA}}
\affiliation[inst2]{organization={Biological \& Environmental Engineering Department},%Department and Organization
            addressline={Cornell University}, 
            city={Ithaca},
            postcode={14853}, 
            state={NY},
            country={USA}}
\fntext[fn1]{Current affiliation: Biological \& Environmental Engineering Department, Cornell University, Ithaca, NY, USA.}           
\affiliation[inst3]{organization={Fruit Research and Extension Center},%Department and Organization
            addressline={The Penn State University}, 
            city={Biglerville},
            postcode={17307}, 
            state={PA},
            country={USA}}
\affiliation[inst4]{organization={School of Business and Technology},%Department and Organization
            addressline={Curry College}, 
            city={Milton},
            postcode={02186}, 
            state={MA},
            country={USA}}            
\affiliation[inst5]{organization={Department of Viticulture and Enology},%Department and Organization
            addressline={Washington State University}, 
            city={Prosser},
            postcode={99350}, 
            state={WA},
            country={USA}}
\affiliation[inst6]{organization={USDA-ARS}, 
            addressline={Horticultural Crops Production and Genetic Improvement Research Unit (HCPGIRU)}, 
            city={Corvallis},
            postcode={973300}, 
            state={OR}, 
            country={USA}}
\fntext[fn2]{Retired, Present address: 1414 Cedar Street, Philomath, OR 97370}
\affiliation[inst7]{organization={Mid-Columbia Agricultural Research and Extension Center},%Department and Organization
             addressline = {Oregon State University}, 
            city={Corvallis},
            postcode={97031}, 
            state={OR},
            country={USA}}
\affiliation[inst8]{organization={Department of Plant Sciences},%Department and Organization
             addressline = {University of Tennessee, Institute of Agriculture}, 
            city={Knoxville},
            postcode={37996}, 
            state={TN},
            country={USA}}
\cortext[cor1]{Corresponding author: Atif Bilal Asad, Email: aba89@cornell.edu}

%% Abstract
\begin{abstract}
%\begin{linenumbers}
Nitrogen (N) is one of the most critical nutrients in winegrape production, influencing vine vigor, fruit composition, and ultimately wine quality. Because soil N availability varies spatially and temporally, accurate estimation of leaf N concentration is essential for optimizing fertilization at the individual plant level. In this study, in-field hyperspectral images ($400$--$1000$nm) were collected from four grapevine cultivars (Chardonnay, Pinot Noir, Concord, and Syrah) across two growth stages (bloom and veraison) during the $2022$ and $2023$ growing seasons at both the leaf and canopy level. An ensemble feature selection framework was developed to identify the most informative spectral bands for N estimation in individual cultivars. The approach effectively reduced redundancy and selected compact, physiologically meaningful combinations of bands spanning the visible, red-edge, and near-infrared regions. At the leaf level, models achieved the highest predictive accuracy (Chardonnay: $R^2$ = 0.82, RMSE = 0.19 \%DW; Pinot Noir: $R^2$ = 0.69, RMSE = 0.20 \%DW). Canopy level predictions also performed well (Chardonnay: $R^2$ =0.65, RMSE =0.17 \%DW; Concord: $R^2$ =0.72, RMSE =0.12\%DW; Syrah: $R^2$ =0.70, RMSE =0.16\%DW). The results from ensemble feature selection showed that white cultivars (e.g., Chardonnay) exhibited balanced contributions from the visible, red edge, and near infrared regions, whereas red cultivars (Pinot Noir, Syrah, and Concord) relied more heavily on visible bands due to anthocyanin–chlorophyll interactions. The analysis demonstrated that leaf level selected N-sensitive bands from the white cultivar (Chardonnay) and the red cultivar (Pinot Noir) were successfully transferred to canopy level for N estimation of their respective cultivar types. The transferred white-cultivar bands improved canopy level prediction accuracy, while the red-cultivar bands maintained comparable performance, confirming that the ensemble feature selection capture robust spectral bands consistent across measurement scales and cultivar. The study demonstrated the potential of using in-field hyperspectral imaging in integrated feature selection and ML techniques to monitor N status in vineyards.
%\end{linenumbers}
\end{abstract}

%% Keywords

\end{frontmatter}

%\linenumbers


\section{Introduction}
\label{intro}
Grape cultivation is a key part of the global agriculture industry. It supports millions of jobs—from vineyard management to winemaking and export logistics, and significantly contributes to economies worldwide, including that of the U.S. The overall grape production in the U.S. is valued at over \$6.8 billion, making it one of the highest-value fruit crops in the country. The industry includes more than $360,000$ hectares of planted grapes, producing about 5.5 million metric tons in 2023 \citep{usda2024}. The profitability of grape cultivation is linked with the effective management of mineral nutrients in vineyards. 
Nitrogen (N) is critical for grapevine physiological processes, influencing growth, productivity, and fruit quality parameters such as sugar content, acidity, phenolics and and aromatic compounds \citep{keller2020science}. Achieving a balance is critical, as excessive and insufficient nitrogen availability can adversely affect grape yield and quality. Abundant availability of N promotes higher canopy growth, resulting in shading of the fruit \citep{smart1991canopy}, which delays fruit ripening and causes poor fruit color development \citep{keller1998interaction}. In addition, excess N resulting from overfertilization remains unused in the soil, which increases the chances of nitrate (NO$_3$) accumulation in natural water bodies if the excess N leaches below the root zone or is lost by run-off \citep{jaynes2001nitrate}. On the other hand, N deficiency can decrease crop yields due to reduced fruit set and berry size, while decreasing yeast assimilable nitrogen (YAN) in the juice \citep{schreiner2013n}, and increasing vulnerability to stress and disease \citep{thomidis2016effects}. \\
Therefore, efficient N management in vineyards requires an accurate assessment of the vine's N status \citep{verdenal2021understanding}. Traditionally, assessing N status in grapevines is based on laborious and time-consuming methods of manually collecting and analyzing leaf tissues through laboratory chemical analysis \citep{robinson}. This labor-intensive process of collecting field samples and conducting subsequent laboratory tests poses a significant obstacle to efficiently monitoring large-scale vineyards. Due to these limitations, current methods of assessing N in vineyards cannot offer the spatial resolution required to represent the variability of N levels within a vineyard, especially for understanding and managing the needs for N in individual vines. Furthermore, there is a time lag between sampling and when the results are available, which hampers real-time monitoring and prompt decision-making.\\
Recent improvements in remote sensing technologies, including hyperspectral and multispectral imaging and ML modeling, provide alternatives for assessing plant nutrient status in a non-destructive and rapid manner with high spatial resolution \citep{paudel2025, chancia2021assessing}. Example studies with these technologies used the spectral signatures of plants in the visible and near infrared regions of the electromagnetic spectrum to identify and estimate plant characteristics associated with nutrient levels in corn \citep{fan2019hyperspectral}, wheat \citep{vigneau2011potential}, barley \citep{xu2014using}, potato \citep{liu2021hyperspectral} and grapevines \citep{peanusaha2024nitrogen,chancia2021assessing,moghimi2018ensemble}.\\
Specifically, hyperspectral imaging (HSI) has emerged as a promising nondestructive technique for assessing plant nutrient status. HSI facilitates the acquisition of high-resolution spectral information across a wide range of wavelengths, spanning from visible to near infrared and shortwave infrared regions of the electromagnetic spectrum \citep{gowen2007hyperspectral}. This high-dimensional spectral data enables detailed characterization of biochemical and physiological properties of plant tissues, as various constituents exhibit unique spectral signatures \citep{kalacska2015estimation}. When coupled with advanced ML algorithms, HSI imaging shows excellent potential for accurately assessing grapevine N levels \citep{chancia2021assessing}. 
HSI-based studies employed various methodologies to estimate N concentration using various ML models. Among these models, some incorporated the entire spectrum range of the sensor. In contrast, others analyzed derivatives of the spectral data \citep{debnath2021identifying, lyu2023assessing}, and several vegetation indexes \citep{peanusaha2024nitrogen}, while other models used spectral transformation techniques \citep{chancia2021assessing}. Some modeling techniques have also used combinations of datasets mentioned above. Specific ML techniques that have been found effective to analyze these complex datasets to determine N levels in plants include support vector machines (SVM) \citep{yao2015evaluation}, random forest regression (RFR) \citep{pullanagari2016mapping}, extreme learning machine (ELM) \citep{peng2022prediction}, and partial least squares regression (PLSR) \citep{chancia2021assessing, peanusaha2024nitrogen}. These ML approaches can learn complex non-linear relationships between the high-dimensional spectral data and corresponding N concentrations, enabling the development of predictive models that can be applied for rapid and non-invasive N monitoring. \\
Unlike multispectral sensors that capture reflectance in a few broad wavelengths, hyperspectral sensors acquire dense spectral information across hundreds of narrow wavelength bands. However, image acquisition in a large number of narrow spectral bands leads to high dimensionality of hyperspectral data, which poses significant challenges to handling huge data volumes during processing, including the need for large memory and computational resources \citep{adao2017hyperspectral}, as well as the potential of data redundancy \citep{bioucas2013hyperspectral}. Specifically, within a large amount of spectral data, only specific wavelengths may have a direct relevance/sensitivity to the target variable, such as varying N status. The rest of the spectral bands may be redundant or irrelevant, which makes interpretation more difficult and increases the likelihood of overfitting the prediction models to specific datasets used in the training process \citep{moghimi2018ensemble}. Another challenge in training ML models with hyperspectral images is the limited availability of data samples, which limits the accuracy and generalizability of the ML models used. It was found that the performance of ML models deteriorates as the number of spectral bands (features) increases while the number of training samples remains fixed. This phenomenon, known as the Hughes effect, leads to a loss of accuracy of the ML model because the model becomes overfitted to the limited dataset used in training the model \citep{ma2013hughes}. To mitigate these challenges and exploit the rich information provided by the hyperspectral data, researchers have explored dimension reduction techniques, such as feature selection methods, to identify the most informative spectral bands for accurate and efficient prediction of plant traits and input status \citep{dong2017dimensionality, rivera2017hyperspectral}.\\
Unlike previous dimensionality-reduction methods that depend on global or linear projections (e.g., local metric learning or statistical latent subspaces), this study introduced an ensemble-based feature-selection framework that first employs hierarchical clustering to reduce spectral multicollinearity and then integrates multiple complementary selectors (Random Frog, SelectKBest, Elastic Net, Random Forest, Lasso, and Ridge) to derive consensus rankings stabilized across growth stages. This unified ensemble strategy preserves physiologically interpretable spectral bands and enhances the robustness of N estimation in grapevine leaves across varying phenological and environmental conditions. Using these selected spectral bands, regression models were trained and evaluated for their ability to predict leaf N concentration across cultivars and measurement scales (Leaf and canopy level). The dataset included four cultivars (Chardonnay, Pinot Noir, Concord, and Syrah) collected during the 2022–2023 growing seasons from two viticultural regions (Washington and Oregon, USA), ensuring environmental and phenological diversity for model robustness. The proposed method not only reduces model complexity but also enhances transferability by selecting bands that remain predictive across growth stages, cultivars, and measurement scales. Finally, transferability was assessed by testing whether spectral bands identified in the white (Chardonnay) and red (Pinot Noir) cultivars at the leaf level remained predictive at the canopy level for respective cultivars. The primary objectives of this study, therefore, were to:

\begin{itemize}
    \item Develop a feature selection method to identify the optimal set of spectral bands for nitrogen assessment in grapevine leaves of individual cultivars.
    \item Develop models that use the optimal set of spectral bands to estimate nitrogen concentration in grapevine leaves of individual cultivars.
    \item Assess the transferability and effectiveness of the selected spectral bands for nitrogen estimation across cultivars and measurement scales (Leaf and canopy level).
\end{itemize}

The results of this work enable rapid and non-destructive assessment of N in vineyards and provide information on N at the vine level. Such data can support the implementation of improved fertilization strategies, such as variable-rate N application, ensuring that each vine receives the optimal amount of N. Furthermore, the identification of a compact set of informative spectral bands offers valuable insight for the future development of customized multispectral sensors optimized for N monitoring in grapevines. These findings suggest a pathway toward simpler, field-deployable sensing systems that capture key physiological signals without the complexity and data volume of full-spectrum hyperspectral imaging.

%% Use \subsection commands to start a subsection.
\section{Materials and Methods}
\label{subsec1}

\subsection{Data Collection} \label{data_collection}
Spectral data and corresponding ground truth measurements of N content (based on leaf blades tissue sampling) were gathered during two phenological stages of grapevine growth, bloom and veraison (i.e., onset of fruit ripening), over two growing seasons of 2022 and 2023. Four grapevine cultivars were used to collect the data listed in Table \ref{tab:data_collection}. This multi-year, multi-cultivar approach facilitated a comprehensive analysis of varying growth stages and environmental conditions.

\begin{table}[H]
  \centering
  \caption{Field data collection efforts over two growing seasons. A tick (\checkmark) indicates successful data collection for the given growth stage; a cross (\texttimes) indicates missing data. Total samples per year are shown in the last two columns, where leaf-level counts refer to the number of leaves and canopy-level counts refer to the number of grapevines.}
  \label{tab:data_collection}
  \resizebox{\textwidth}{!}{%
    \begin{tabular}{|c|c|c|c|c|c|c|c|}
      \hline
      \multirow{2}{*}{Cultivar} & \multirow{2}{*}{Leaf/Canopy Level} & \multicolumn{2}{c|}{2022} & \multicolumn{2}{c|}{2023} & \multicolumn{2}{c|}{Total Samples} \\
      \cline{3-8}
       & & Bloom & Veraison & Bloom & Veraison & 2022 & 2023 \\
      \hline
      Chardonnay & Leaf   & \checkmark & \checkmark & \checkmark & \checkmark & 113 & 102 \\
      \hline
      Pinot Noir & Leaf   & \checkmark & \checkmark & \checkmark & \checkmark & 169 & 108 \\
      \hline
      Chardonnay & Canopy & \checkmark & \checkmark & \checkmark & \checkmark & 113 & 109 \\
      \hline
      Syrah      & Canopy & \texttimes & \checkmark & \checkmark & \checkmark & 100  & 137 \\
      \hline
      Concord    & Canopy & \checkmark & \checkmark & \checkmark & \checkmark & 75  & 66 \\
      \hline
    \end{tabular}%
  }
\end{table}

\subsubsection{Leaf Level Spectral and Leaf Tissue Sampling} \label{leaf_level_Spectral_and_tissue sampling}
This study was conducted in two commercial vineyard blocks located in Oregon, USA; Vitis vinifera L. cv. Pinot Noir vineyard (44°53'16" N; 122°51'06" W) and Chardonnay vineyard (46°17'49" N; 119°44'07" W). The vines were selected randomly from each vineyard for data collection, ensuring a diverse sampling. From each data vine, two leaves, well-exposed to the sun, healthy, fully developed, and proximal to the grape clusters on the same side of the canopy, were tagged for sampling. For the spectral data acquisition, we employed a VNIR hyperspectral camera (Nano-Hyperspec®, Headwall Photonics, Bolton, MA, USA) mounted on a ground-based tripod. This setup allowed precise positioning of the camera—approximately $2$ meters from the target vine and $1$ meter above the ground for optimal focus on the middle of the vine's vertical canopy. The camera's line scanning capability covered a spectral range from $400$ to $1000nm$ across $274$ spectral bands. The operational parameters of the camera were tuned to capture high-resolution images $(\sim 3200 \times 640\,\text{pixels})$. The exposure times were managed between $9$ and $15$ milliseconds to maintain above $90\%$ saturation level, focusing on a white reference to ensure consistent image quality. Concurrently, the imaging frame period was synchronized with the exposure time, facilitating the determination of the scanning rate. The angular scanning speed, which varied between 1.20 and 1.85°/s, was calculated by using the FOV Calculator from HyperSpec III software, developed by Headwall Photonics \citep{kang2023estimating}. Each image acquired by the camera included both tagged leaves within the field of view, as well as both white and dark reference surfaces for calibration. Following the imaging of each vine, the leaf blades were carefully removed from the vines and individually sealed in paper bags, then transported to the laboratory for subsequent nutrient analysis. 

\subsubsection{Measured Nitrogen at Leaf Level} \label{measured_N}
\label{TN_LL}
\begin{figure}[!tbp]
    \centering
    \begin{subfigure}[t]{0.48\textwidth}
        \centering
        \includegraphics[width=\textwidth]{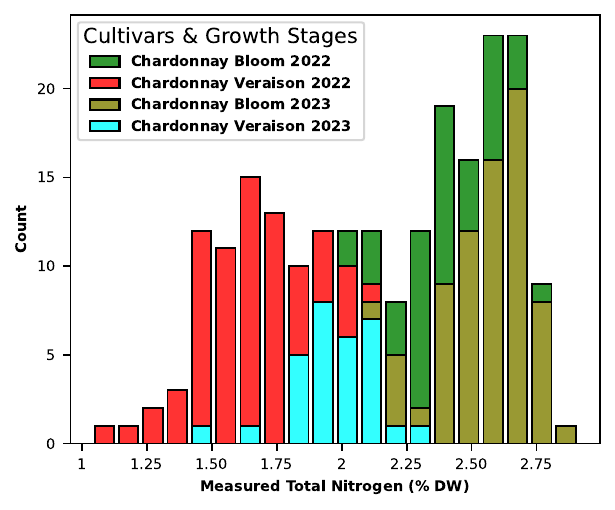}
        \subcaption{Chardonnay}
        \label{fig:histogram_canopy}
    \end{subfigure}
    \hfill
    \begin{subfigure}[t]{0.48\textwidth}
        \centering
        \includegraphics[width=\textwidth]{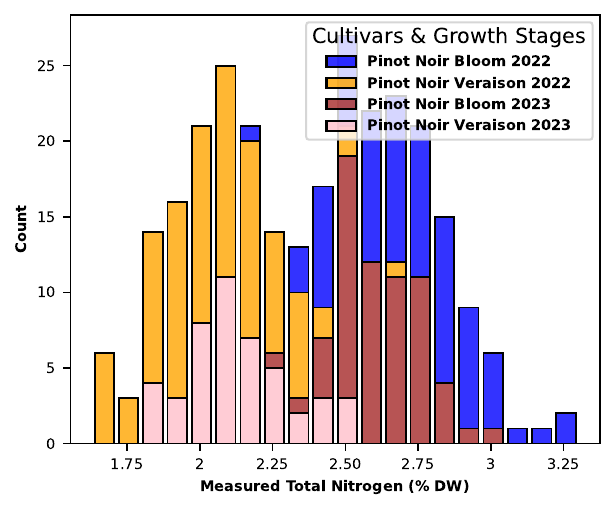}
        \subcaption{Pinot Noir}
        \label{fig:histogram_leaf}
    \end{subfigure}
    \caption{Distribution of measured N concentration in grapevine leaves at leaf level (a) Chardonnay; (b) Pinot Noir.}
    \label{fig:TN_LL_histo}
\end{figure}

The summary of N concentrations of the grapevine leaves at leaf level measured using chemical analysis presented in Table \ref{tab:TN_LL}. The N concentration exhibited significant variability among cultivars and seasons, as shown in Figure \ref{fig:TN_LL_histo}. In leaf level, the N concentration of both Chardonnay and Pinot Noir cultivars typically varied from $1.04\%$ to $3.30\%$ dry weight (DW) of grapevine leaves. The histograms demonstrate the variations in N concentration in grapevine leaves throughout two growing seasons, highlighting the impact of phenological stage on leaf N concentration.

\begin{table}[ht]
\centering
\caption{Descriptive statistics of N concentration per unit dry weight (DW) of grapevine leaves at leaf level, summarized by cultivar.}
\resizebox{\textwidth}{!}{%
\begin{tabular}{|l|l|c|c|c|}
\hline
\textbf{Measurement Level} & \textbf{Cultivar} & \textbf{Min–Max (\% DW)} & \textbf{Mean (\% DW)} & \textbf{STD (\% DW)} \\
\hline
\multirow{2}{*}{Leaf Level} 
& Chardonnay   & 1.04–2.91 & 2.14 & 0.43 \\
& Pinot Noir   & 1.63–3.30 & 2.38 & 0.35 \\
\hline
\end{tabular}%
}
\label{tab:TN_LL}
\end{table}

\subsubsection{Canopy Level Spectral and Leaf Tissue Sampling}
\label{canopy_level_Spectral_and_tissue_sampling}
This data collection was conducted with three grapevine cultivars located in Washington, USA, two commercial wine grape blocks at Columbia Crest Winery (45°57’25.95” N, 119°36’34.41” W) of V. vinifera cv. Chardonnay and Syrah and a block (46°19’4.84” N, 119°55’56.51” W) of Concord, a juice grape hybrid cultivar with Vitis labrusca L. and V. vinifera ancestry. To maintain uniformity in data collection, the selected data vines were consistently sampled at both growth stages and both growing seasons listed in Table \ref{tab:TN_CL}. The image acquisition process was carried out exactly as described in Section \ref{leaf_level_Spectral_and_tissue sampling}. However, leaf blade tissue sampling was conducted differently. For this canopy level data, each tissue sample contained 20 to 40 completely developed leaves carefully removed from each data vine near the grape clusters. To maintain the biochemical integrity of the gathered leaves, they were sealed in a plastic bag before being sent to a commercial laboratory (Soiltest Farm Consultants, Moses Lake, WA, USA) for nutrient analysis. 

\begin{table}[ht]
\centering
\caption{Descriptive statistics of N concentration per unit dry weight (DW) of grapevine leaves, summarized by cultivar and measurement level.}
\label{tab:TN_CL}
\resizebox{\textwidth}{!}{%
\begin{tabular}{|l|l|c|c|c|}
\hline
\textbf{Measurement Level} & \textbf{Cultivar} & \textbf{Min–Max (\% DW)} & \textbf{Mean (\% DW)} & \textbf{STD (\% DW)} \\
\hline
\multirow{3}{*}{Canopy Level} 
& Chardonnay   & 2.38–3.70 & 2.91 & 0.28 \\
& Concord      & 2.62–3.55 & 3.12 & 0.22 \\
& Syrah        & 2.36–3.72 & 3.03 & 0.30 \\
\hline
\end{tabular}%
}
\end{table}

\subsubsection{Measured Nitrogen at Canopy Level}
\label{TN_CL}
\begin{figure}[!tbp]
    \centering
    \begin{subfigure}[t]{0.48\textwidth}
        \centering
        \includegraphics[width=\textwidth]{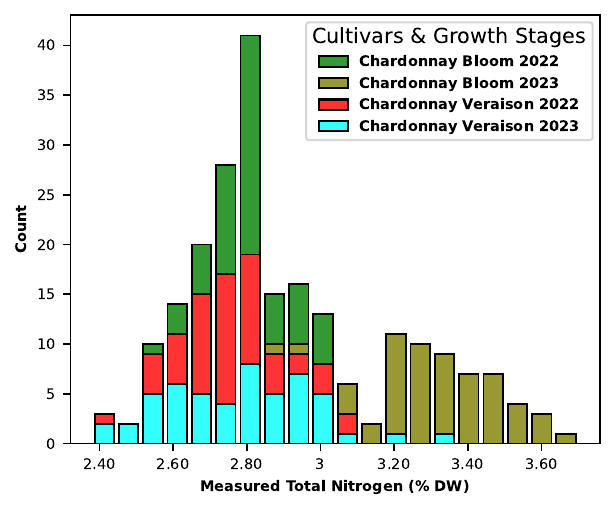}
        \subcaption{Chardonnay}
    \end{subfigure}
    \hfill
    \begin{subfigure}[t]{0.48\textwidth}
        \centering
        \includegraphics[width=\textwidth]{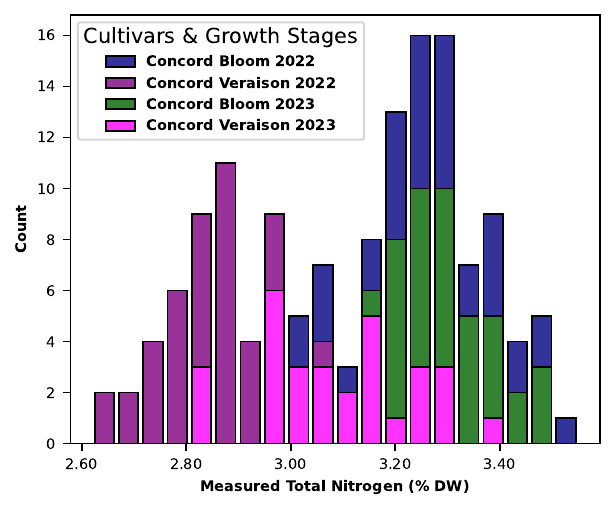}
        \subcaption{Concord}
    \end{subfigure}
    \hfill
    \begin{subfigure}[t]{0.48\textwidth}
        \centering
        \includegraphics[width=\textwidth]{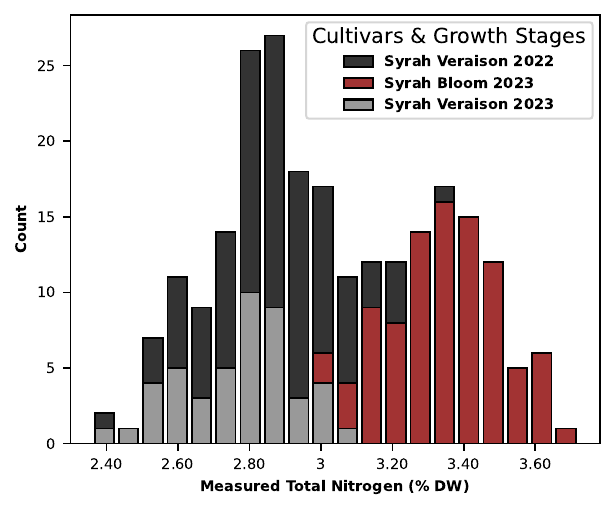}
        \subcaption{Syrah}
    \end{subfigure}
    \caption{Distribution of measured N concentration in grapevine leaves at leaf level (a) Chardonnay; (b) Concord; (c) Syrah.}
    \label{fig:TN_CL_histo}
\end{figure}

The summary of N concentrations of the grapevine leaves measured at canopy level using chemical analysis presented in Table \ref{tab:TN_CL}. In contrast to leaf level, the canopy level N data include three cultivars, Chardonnay, Syrah, and Concord, which have N concentration primarily ranging from $2.36\%$ to $3.72\%$ DW of leaves, with a mean value of approximately 2.9\% DW. The N concentration exhibited significant variability among cultivars and seasons, as shown in Figure \ref{fig:TN_CL_histo}.

From the histogram, it was observed that the N concentration was consistently higher during bloom (peak to the right) compared to the same during veraison (peak to the left) across all cultivars except Chardonnay in 2022, as shown in Figure \ref{fig:TN_CL_histo}. This difference primarily reflects the change in leaf age during the growing season: N concentrations are typically high in young tissues and decrease as leaves age, along with a decline in chlorophyll content \citep{keller2020science, verdenal2021understanding}. It was also observed that canopy level samples displayed a narrower range of N concentrations but higher mean values compared to leaf level samples. This variation is largely attributed to the sampling method at the canopy level, which involves collecting multiple leaves from different parts of the vine, capturing a broader picture of the overall N concentration of the vine. This method removes individual leaf level variability caused by localized environmental factors, such as shading or exposure to sunlight. Additionally, looking at just the Chardonnay data, it is also likely that the Oregon vineyards had overall lower N status than the Washington vineyards shown in Figures \ref{fig:TN_LL_histo} and \ref{fig:TN_CL_histo}. Furthermore, each vineyard site for data collection received varying amounts of fertilizer applications. These differences in fertilization and variations in environmental conditions, soil types, and management practices between vineyards significantly influence N concentration in leaves.

\subsection{Data Analysis} \label{data_analysis}
\subsubsection{Spectral Data Preparation - Leaf Level} \label{Spectral_Data_Preparation_Leaf_level}
At the leaf level, the N concentration was measured from individual leaf, and the corresponding spectral data was extracted for each leaf. In contrast, canopy level processing aggregates typically 20–40 leaves per vine to produce a composite measurement representing the overall N concentration of the canopy, which offers a broad view of the nutritional status of the vine, while for spectral data, average spectra of all the leaf pixels of the canopy were extracted as discussed in Section \ref{Spectral_Data_Preparation_Canopy_level}. The process of extracting leaf level spectral data consisted of reflectance correction, leaf segmentation, and reflectance spectra extraction from individual leaf areas. First, radiometric calibration was performed to convert raw digital numbers (DN) to reflectance values using Equation \ref{eq:corrected_intensity}:

\begin{equation}
    \label{eq:corrected_intensity}
    I_{c(i,j,\lambda)} = \frac{I_{r(i,j,\lambda)} - I_{\text{dark}(i,j,\lambda)}}{I_{\text{white}(i,j,\lambda)} - I_{\text{dark}(i,j,\lambda)}} \times (\text{Absolute reflectance of White Reference})
\end{equation}

where, $I_{c}$ is the corrected reflectance of a given pixel, $I_{r}$ is the intensity of light at the pixel; $i$ and ${j}$ are the spatial coordinates of the pixel,  ${\lambda}$ is the wavelength; and $I_{\text{white}}, I_{\text{dark}}$ are the white and dark references \citep{kang2023estimating}. The absolute reflectance of white reference is the known standardized reflectance value of a white reference material (often close to 1, or 100 \% reflectance) under ideal conditions. In Eq. \ref{eq:corrected_intensity}, the term  $(I_r - I_{\text{dark}}) / (I_{\text{white}} - I_{\text{dark}})$ calculates the relative reflectance of the target compared to the white reference under the same lighting conditions; multiplying this relative reflectance by the "Absolute reflectance of White Reference" converts this relative measurement into an absolute reflectance value of the target pixel $I_{c}$.
\begin{figure}[!ht]
    \centering
    \includegraphics[width=0.5\textwidth]{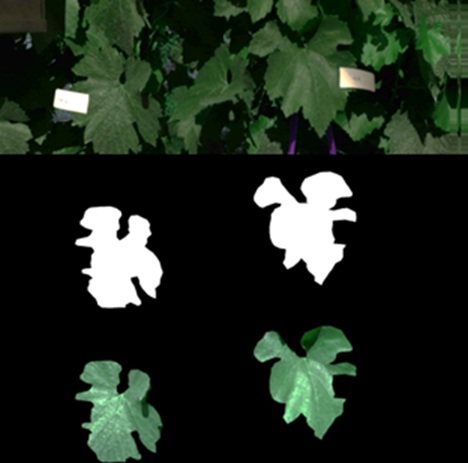}
    \caption{Sample leaves used in the study were labeled using white tags (top); binary masks of the tagged leaves were created (middle), which were then used to extract the sample leaf areas (bottom).}
    \label{fig:RGB_Segmented_leaves}
\end{figure}
The RGB images were generated using the wavelength bands $487.642$ nm (B), $547.321$ nm (G), and $677.732$ nm (R) to facilitate accurate segmentation of the leaf area from the hyperspectral data. These bands were chosen through trial and error as they provided a good visual representation that aids in identifying the tagged leaf. After reflectance correction, the leaves were manually segmented to accurately delineate the complex canopy structure, which was chosen because of the challenge due to the similarity in color and texture between the leaves and background, as shown in Figure \ref{fig:RGB_Segmented_leaves}. Once leaves were segmented, the spectral data of all pixels within each leaf area were averaged to obtain a single observation that represents the sample leaf.

\begin{figure}[!ht]
    \centering
    \includegraphics[width=0.9\textwidth]{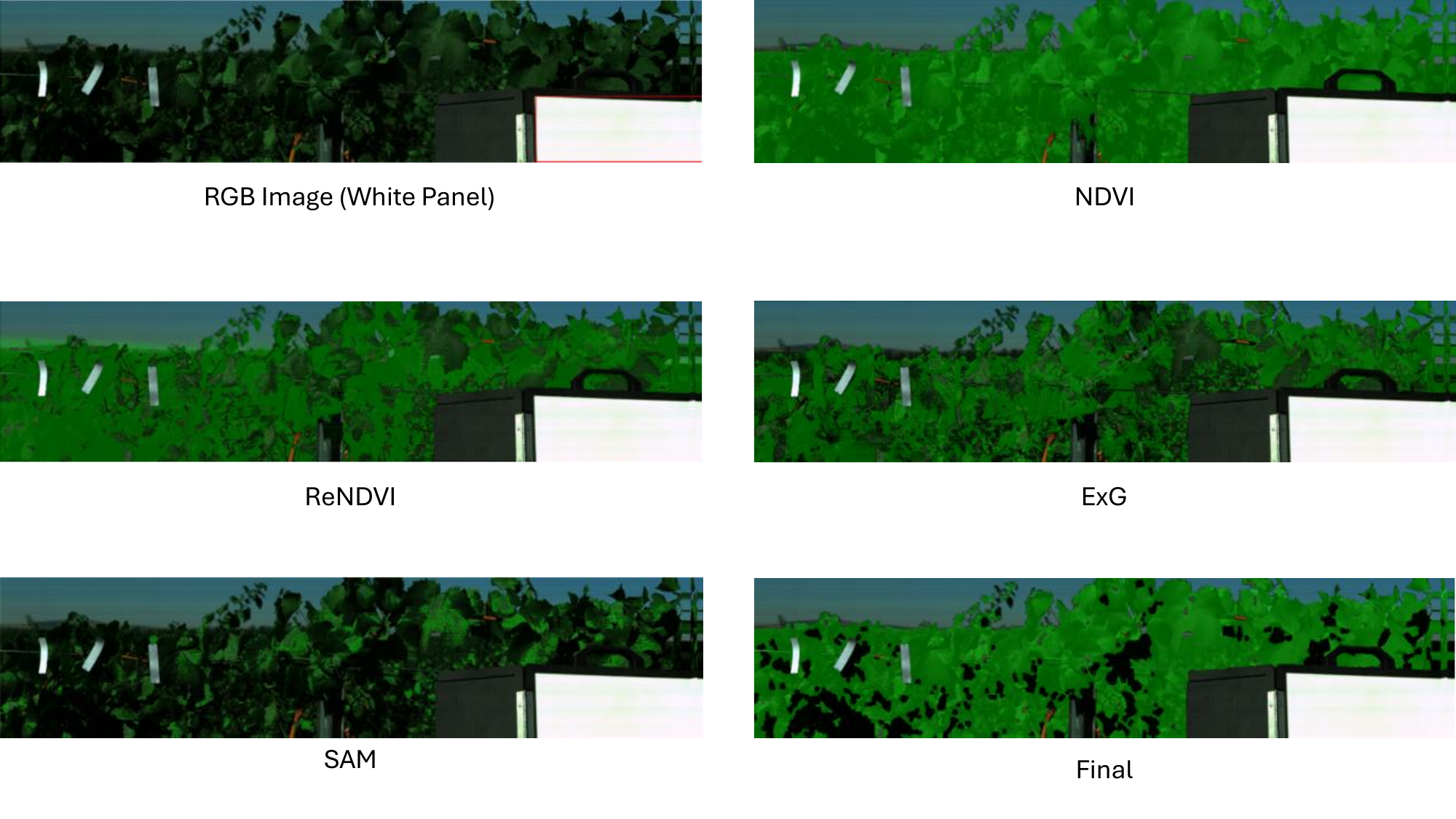}
    \caption{Leaf pixels segmentation applied to grapevine canopy hyperspectral image. The RGB+Panel image (top-left) shows the original scene with the white reference panel outlined in red. The subsequent panels show vegetation masks (green overlay) derived from NDVI, NDRE, ExG, and SAM indices, with the final panel presenting the combined leaf mask that integrates all four approaches to segment leaf pixels.}
    \label{fig:CL_Leaf_Pixel_Segmentation}
\end{figure}

\subsubsection{Spectral Data Preparation - Canopy Level} \label{Spectral_Data_Preparation_Canopy_level}
At the canopy level, measurements were made to capture the overall N status of entire vine canopies rather than focusing on individual leaves. This approach is essential because the chemical analysis of tissue samples by commercial testing laboratories is typically performed on a composite sample, obtained by mixing $20$ to $40$ leaves from a single vine, which represents the average concentration of nutrients throughout the canopy. To represent this composite measurement approach using spectral data, after performing the spectral correction procedure outlined in Section \ref{Spectral_Data_Preparation_Leaf_level}, leaf pixels were automatically segmented using a combination of vegetation indices and spectral mapping techniques as shown in Figure \ref{fig:CL_Leaf_Pixel_Segmentation}. The indices employed were the Normalized Difference Vegetation Index (NDVI), the Normalized Difference Red Edge Index (NDRE), and the Excess Green Index (ExG), along with Spectral Angle Mapping (SAM). Thresholds for each index were adaptively determined using Otsu's method within vegetation seed regions to account for varying illumination and seasonal conditions. The final leaf mask retained pixels meeting ExG, combined NDVI and NDRE thresholds, or high SAM similarity. After morphological filtering and removal of components smaller than 500 pixels, mean reflectance spectra were extracted from all pixels in the masked regions.

\subsection{Pre-Treatment of Data} \label{spectral_preprocessing}
The data preprocessing and analysis methodology described in this and the following sections were applied separately to each cultivar. Furthermore, even within a single cultivar like Chardonnay, data from different measurement scales (e.g., leaf and canopy level) were processed as independent datasets. The preprocessing steps included edge band removal, outlier removal, and the application of the standard normal variate (SNV) and Savitzky-Golay (SG) filter for smoothing. It was observed that the first and last $5$ bands of the hyperspectral images exhibited significant distortions, likely due to sensor noise or edge effects inherent to the spectral data acquisition process. Outliers were removed from spectral data and measured N data using the Median Absolute Deviation (MAD) method applied to each unique combination of cultivar, measurement scale (leaf level, canopy level), growth stage, and year. MAD was selected for outlier detection because data exploration revealed extreme values in both N concentrations and spectral data. Since MAD uses the median rather than the mean, it remains unaffected by these extreme values, providing reliable outlier identification. The MAD for each category was calculated using Equation \ref{eq:MAD}:
\begin{equation} \label{eq:MAD}
\text{MAD} = \text{median}(|X_i - \tilde{X}|)
\end{equation}
\noindent where $X_i$ represents individual observations and $\tilde{X}$ is the median of the dataset within each category. Samples were classified as outliers when their deviation exceeded the threshold according to Equation \ref{eq:outlier}:
\begin{equation} \label{eq:outlier}
|X_i - \tilde{X}| > \tau \times \text{MAD}
\end{equation}
\noindent where $\tau = 4.5$ was the selected threshold value. This criterion was applied independently to both measured N concentration and spectral reflectance values, with samples flagged as outliers if they exceeded the threshold in either domain. The measure N concentrations in Section \ref{TN_CL} and \ref{TN_LL} are after outlier removal. These outliers and distortions can lead to inaccuracies in subsequent analysis, such as feature selection and modeling. To ensure the integrity and reliability of the analysis, these outliers and edge spectral bands were excluded from the datasets. Following outlier removal, SNV transformation was applied to minimize multiplicative scatter effects and correct for baseline variations caused by differences in light scattering \citep{rinnan2009review}. Finally, a SG filter was used for smoothing spectral noise while preserving important spectral features and peak shapes \citep{rinnan2009review}. The optimal window length for the SG filter was determined using signal-to-noise ratio (SNR) considerations. This comprehensive data preprocessing approach ensured effective reduction of noises in the spectral data while preserving important spectral features. 

\subsection{Spectral Feature Selection}
\subsubsection{Hierarchical Clustering for Redundant Feature Elimination}
\label{hierarchical_clustering}
The high dimensionality of the hyperspectral data often results in significant redundancy among the spectral bands, which can complicate the feature selection process, increase computational demands, and potentially lead to model overfitting. To improve the efficiency and effectiveness of the feature selection process, redundant spectral bands were removed from the datasets using hierarchical clustering. For this clustering approach, the correlation-based distance matrix of the spectral bands was calculated to understand the relationships between different spectral bands as shown in Equation \ref{distance_matrix},  where $\rho$($i$,$j$) represents the Pearson correlation coefficient between spectral bands $i$ and $j$.

\begin{equation}
d(i,j) = 1 - |\rho(i,j)|
\label{distance_matrix}
\end{equation}

The absolute correlation values ensured that both positive and negative correlations were treated as meaningful relationships. Then, hierarchical clustering was performed on the distance matrix using the complete linkage clustering method \citep{sharma2019comparative}. In this method, the distance between two clusters is defined as the maximum distance between any pair of features within the clusters. This means that clusters are merged only when all of the members are relatively similar, resulting in compact and well-separated groups \citep{sharma2019comparative}. Optimal clustering thresholds were determined through automated silhouette analysis \citep{lai2025silhouette}, twenty candidate thresholds, evenly spaced from 1\% to 95\% of the maximum linkage distance, were evaluated. The threshold yielding the highest silhouette coefficient was selected. The silhouette coefficient was calculated using Equation \ref{eq:silhouette}:

\begin{equation}
s(i) = \frac{b(i) - a(i)}{\max(a(i), b(i))}
\label{eq:silhouette}
\end{equation}

\noindent where $a(i)$ represents the mean intra-cluster distance for spectral band $i$, and $b(i)$ represents the mean distance of the spectral band $i$ to the nearest neighboring cluster. Silhouette values range from $-1$ to $+1$, with higher values indicating better cluster assignment. To ensure practical applicability, the final number of clusters was restricted between $10$ and $30$, balancing the dimension reduction with the preservation of spectral resolution.
The representative spectral band of each cluster was selected with a consensus score(CS) that integrated three evaluation criteria. The first criterion is the normalized absolute Pearson correlation coefficient between the spectral bands and N, which quantifies the linear relationship between the two. To capture non-linear dependencies, we used Mutual Information(MI) \citep{ross2014mutual}; specifically the normalized MI score between the spectral bands and N. Finally, the centrality component assessed a spectral band's representativeness by measuring its position relative to the cluster centroid in correlation space, as estimated by the Equation \ref{eq: centrality}, where $\bar{\rho}_{\text{cluster}}(\lambda)$ is the mean correlation of spectral band $\lambda$ with all bands in its cluster, and $\bar{\rho}_{\text{centroid}}$ is the cluster centroid value.

\begin{equation}
\text{Centrality}(\lambda) = 1 - \frac{|\bar{\rho}_{\text{cluster}}(\lambda) - \bar{\rho}_{\text{centroid}}|}{\text{Max\ Distance\ in\ Cluster}}
\label{eq: centrality}
\end{equation}

For CS, each component of CS was first normalized to a $0-1$ scale. The three normalized components were then combined into the final CS using a weighted summation given by Equation \ref{eq:consensus_score}.
\begin{equation}
\mathrm{Consensus\ Score}(\lambda) = w_1 \times \mathrm{Correlation}_{\mathrm{norm}}(\lambda) + w_2 \times \mathrm{MI}_{\mathrm{norm}}(\lambda) + w_3 \times \mathrm{Centrality}_{\mathrm{norm}}(\lambda)
\label{eq:consensus_score}
\end{equation}

\noindent where $\lambda$ denotes the spectral band and the weights are defined as the vector $\mathbf{w} = [w_1, w_2, w_3] = [0.5, 0.4, 0.1]$ for correlation ($\text{Correlation}_{\text{norm}}$), mutual information ($\text{MI}_{\text{norm}}$), and centrality component ($\text{Centrality}_{\text{norm}}$), respectively, subject to the normalization constraint $\sum_{i=1}^{3} w_i = 1$.
For each cluster, the spectral band that achieved the highest weighted CS was selected as the representative spectral band. This identified the specific spectral band within the cluster that achieved the highest consensus score by ensuring that the most informative and representative spectral band is selected from each cluster. This robust approach of selecting representative spectral bands ensured that the final set of spectral bands offered both high relevancy to N and strong spectral neighborhood representation across all identified clusters.

\subsubsection{Ensemble Feature Selection Method} \label{ensemble}
An ensemble feature selection technique was utilized to identify spectral bands for predicting N concentration in grapevine leaves. We implemented an ensemble feature selection methodology designed to improve robustness and feature stability through a ranking aggregation. This approach systematically combined eight distinct ML models, each capturing different facets of feature relevance, ensuring that the selected spectral bands were consistently informative across multiple ML model's frameworks. These included: (1) a filter-based method: SelectKBest \citep{pedregosa2011scikit} that evaluates features using univariate statistical tests; (2) regularization-based methods including Lasso \citep{tibshirani1996regression}, Ridge \citep{paul2016feature}, and Elastic Net \citep{zou2005regularization} that identify features through coefficient penalization; (3) tree-based ensemble methods includes Random Forest \citep{chancia2021assessing}, Extra Trees \citep{alfian2022predicting}, and Gradient Boosting \citep{adler2022feature} that capture nonlinear relationships and interactions; and (4) a metaheuristic Random Frog algorithm \citep{li2012random} that explores feature space through stochastic optimization. This methodological diversity ensures that selected spectral bands are consistently informative regardless of the underlying algorithmic assumptions. The Random Frog algorithm \citep{li2012random} was customize to implement as an optimization technique. This maintained multiple feature subsets (frogs) that evolved through iterative selection, crossover, and mutation. The algorithm was initialized with 20 feature subsets, each containing 10 randomly selected bands. In each iteration, the performance $P_j^{(t)}$ of every frog $j$ was evaluated using cross-validated $R^2$ scores, as shown in Equation \ref{eq:frog_performance}, against the target variable vector ($\mathbf{y}$).

\begin{equation}
P_j^{(t)} = \text{CV-}R^2(\mathbf{X}_{F_j}, \mathbf{y})
\label{eq:frog_performance}
\end{equation}

\noindent where $\mathbf{X}_{F_j}$ represents the feature matrix containing only the spectral bands selected by frog $j$, and $\mathbf{y}$ is the N vector. The top-performing frogs are retained based on the elite selection criterion defined in Equation \ref{eq:elite_selection}:

\begin{equation}
\mathcal{S}_{\text{elite}} = \{j : P_j^{(t)} \geq P_{(N_{\text{frogs}}/2)}^{(t)}\}
\label{eq:elite_selection}
\end{equation}

\noindent where $P_{(N_{\text{frogs}}/2)}^{(t)}$ represents the median performance threshold. New offspring are generated through crossover operations between elite frogs, followed by random mutation to maintain population diversity. The bootstrap iterations ($i=100$) were performed using stratified sampling based on growth stage combinations to ensure proportional representation of both growth stages. This approach maintains the original distribution of growth stage combinations while providing diverse training subsets for robust feature evaluation. For each feature selection method $m$ and iteration $i$, importance scores were converted to rankings using the transformation shown in Equation \ref{eq:ranking_conversion}:

\begin{equation}
R_m^{(i)}(\lambda) = \text{rank}(-I_m^{(i)}(\lambda))
\label{eq:ranking_conversion}
\end{equation}

\noindent where $I_m^{(i)}(\lambda)$ represents the importance score assigned to wavelength $\lambda$ by method $m$ in iteration $i$, and the negative sign ensures that higher importance corresponds to better (lower) ranks. The aggregated ranking across all methods for each iteration was calculated using the mean ranking formula in Equation \ref{eq:mean_ranking}:

\begin{equation}
\bar{R}^{(i)}(\lambda) = \frac{1}{M} \sum_{m=1}^{M} R_m^{(i)}(\lambda)
\label{eq:mean_ranking}
\end{equation}

\noindent where $M = 8$ represents the total number of feature selection methods. Feature stability was quantified by tracking the frequency of appearance in top-$k$ selections across iterations. For each wavelength $\lambda$, the stability score was computed using Equation \ref{eq:stability_score}:

\begin{equation}
S_{\text{stability}}(\lambda) = \frac{1}{n_{\text{iter}}} \sum_{i=1}^{n_{\text{iter}}} \mathbb{I}(\bar{R}^{(i)}(\lambda) \leq k)
\label{eq:stability_score}
\end{equation}

\noindent where $\mathbb{I}(\cdot)$ is the indicator function, $k = 10$ represents the top-$k$ threshold, and $n_{\text{iter}}$ is the number of bootstrap iterations. This metric captures the consistency with which features are selected across different data subsets and algorithmic perspectives. The final feature scores integrated both ranking performance and selection stability through the standardized combination presented in Equation \ref{eq:final_score}:

\begin{equation}
S_{\text{final}}(\lambda) = \alpha \cdot z_{\text{stability}}(\lambda) + (1-\alpha) \cdot z_{\text{ranking}}(\lambda)
\label{eq:final_score}
\end{equation}

\noindent where $z_{\text{stability}}(\lambda)$ and $z_{\text{ranking}}(\lambda)$ represent the standardized z-scores of stability and mean ranking metrics, respectively, and $\alpha = 0.4$ is the stability weight parameter. The z-score standardization was performed according to Equation \ref{eq:zscore}:

\begin{equation}
z_{\text{metric}}(\lambda) = \frac{\text{metric}(\lambda) - \mu_{\text{metric}}}{\sigma_{\text{metric}}}
\label{eq:zscore}
\end{equation}

\noindent where $\mu_{\text{metric}}$ and $\sigma_{\text{metric}}$ represent the mean and standard deviation of the metric across all spectral band $\lambda$. To ensure spectral diversity and avoid selection of redundant neighboring wavelengths, a minimum spacing constraint was enforced during final feature selection. The spectral bands were selected in descending order of $S_{\text{final}}(\lambda)$ scores, subject to the spatial constraint defined in Equation \ref{eq:spacing_constraint}, which ensured that in the final selection each spectral band exhibited a high $S_{\text{final}}(\lambda)$ score while maintaining a minimum spectral distance of $d_{\text{min}} = 10nm$:

\begin{equation}
\min_{j \in \mathcal{F}_{\text{selected}}} |\lambda_i - \lambda_j| \geq d_{\text{min}}
\label{eq:spacing_constraint}
\end{equation}

\noindent where $\mathcal{F}_{\text{selected}}$ represents the set of previously selected spectral band, $\lambda_i$ is the candidate band, and $d_{\text{min}}$ is the minimum allowable spacing. The ensemble approach thus provided a robust framework for feature selection that balanced predictive relevance, selection stability, and spectral diversity, resulting in a reduced spectral bandset optimized for N concentration prediction.

\subsubsection{Sequential Feature Optimization via PLSR} \label{plsr_opt}
Following the selection of the ensemble features, a systematic optimization process was implemented to determine the spectral bands required to achieve optimal accuracy for the prediction of N concentration in grapevine leaves using partial least squares regression (PLSR). The optimization strategy addressed a fundamental question: whether more spectral bands can potentially capture additional spectral information or introduce noise and computational complexity that degrade model performance. To resolve this trade-off, features were systematically added in order of decreasing importance, and model performance was evaluated at each step to identify the optimal spectral band subset size. The process began by ranking all ensemble-selected bands in descending order of their $S_{\text{final}}(\lambda)$, ensuring that the most informative and stable spectral bands were prioritized for inclusion. Starting with the single highest-ranked spectral band, additional bands were sequentially incorporated to create nested band subsets of increasing dimensionality. For each spectral band subset evaluation, spectral bands were combined with categorical growth stage information to create comprehensive input matrices as described in Equation \ref{eq:feature_matrix}:

\begin{equation}
\mathbf{X}_k = [\mathbf{X}_{\text{spectral}}^{(k)} \mid \mathbf{X}_{\text{growth stage}}]
\label{eq:feature_matrix}
\end{equation}

\noindent where $\mathbf{X}_{\text{spectral}}^{(k)}$ contains the top $k$ ensemble-selected spectral bands, and $\mathbf{X}_{\text{growth stage}}$ represents one-hot encoded growth stage variables that capture phenological effects on N and spectral relationships. A critical component of the optimization process involved determining the appropriate number of Partial Least Square (PLS) components for each feature subset size. Using too few components may underfit the model, while excessive components can lead to overfitting, particularly with smaller feature sets. The optimization of the PLS component followed a three-tier constraint. First, an upper bound of 20 components ensured computational efficiency. Second, the number of components could not exceed the number of input features due to the mathematical limitations of PLS decomposition. Third, sufficient samples were reserved for error estimation by limiting components to the sample size minus 2. The most restrictive of these three constraints determined the maximum number of components. Cross-validation employed the GroupKFold methodology, using growth stage-year combinations as grouping variables, to ensure robust performance estimation. This approach prevented data leakage by maintaining all samples from the same experimental condition within the same fold during both training and validation splits. The number of cross-validation folds corresponded to the number of available growth stage-year combinations within each cultivar dataset. For each combination of feature subset size and component number, model performance was evaluated through cross-validated predictions. The optimal number of components for each feature subset was selected by maximizing the cross-validated coefficient of determination, ensuring that each feature subset was evaluated at its maximum potential performance level for fair comparison across different spectral band subsets. The sequential evaluation process systematically tested all feature subset sizes from $1$ to the total number of features selected by the ensemble feature selection approach. For each subset size, both the coefficient of determination ($R^2$) and the root mean squared error (RMSE) were recorded to assess the performance of the model. The globally optimal subset size of the features was determined using Equation \ref{eq:optimal_features}:

\begin{equation}
k_{\text{opt}} = \underset{k}{\arg\max} \, {(R^2_{\text{cv}})}^{(k)}
\label{eq:optimal_features}
\end{equation}

\noindent where ${R^2_{\text{cv}}}^{\,(k)}$ represents the cross-validated R² achieved by the optimal PLSR configuration using $k$ features. This approach provided an objective framework for identifying the set of spectral bands to achieve optimal N concentration prediction. The methodology forged a balance between model complexity and predictive performance, while maintaining the spectral diversity established through ensemble feature selection, yielding a set of spectral bands optimized for both predictive accuracy and model simplicity.

\subsection{Machine Learning Model Development}
The sequential PLSR optimization identified the optimal subset of spectral bands for N concentration prediction. While PLSR provided a computationally efficient approach for determining optimal feature subset size through iterative evaluation, it assumes linear relationships between spectral reflectance and N concentration. To capture potential nonlinear spectral-N relationships, four machine learning algorithms representing distinct modeling paradigms were subsequently evaluated: Elastic Net (EN), Support Vector Regression with RBF kernel (SVR-RBF), XGBoost, and Gaussian Process Regression (GPR). These machine learning models were trained on raw reflectance values to validate whether the ensemble-selected spectral bands maintain predictive accuracy without preprocessing, and to enhance practical field deployment. Input features consisted of the PLSR-optimized spectral bands raw reflectance vlaues and growth stage (Bloom or Veraison) for each cultivar. Growth stage was one-hot encoded, and spectral features were standardized using z-score normalization (mean = 0, standard deviation = 1) to ensure equal feature contribution during training.

Model hyperparameters were optimized using a robust nested cross-validation approach. In the outer loop, a 5-fold stratified cross-validation, repeated twice (resulting in 10 train-test splits), was used to evaluate performance, ensuring balanced representation of bloom and veraison samples across the folds. In the inner loop, a 5-fold stratified cross-validation was applied for hyperparameter tuning. Randomized search with 200 iterations was employed for all the models: EN, SVR-RBF, XGBoost, and GPR. This procedure ensured unbiased performance estimation and reduced the risk of overfitting.

Model performance was assessed using $R^2$ and RMSE. $R^2$ quantified the proportion of variance in N concentration explained by the model using Equation \ref{rsquare}:

\begin{equation}
\label{rsquare}
R^2 = 1 - \frac{\sum_{i=1}^{n}(y_i - \hat{y}_i)^2}{\sum_{i=1}^{n}(y_i - \bar{y})^2}
\end{equation}

where $y_i$ represents the measured N concentration, $\hat{y}_i$ the predicted value, $\bar{y}$ the mean of measured values, and $n$ the number of samples. RMSE measured the average prediction error in the same units as the target variable (\%DW) with Equation \ref{rmse}:

\begin{equation}
\label{rmse}
RMSE = \sqrt{\frac{1}{n}\sum_{i=1}^{n}(y_i - \hat{y}_i)^2}
\end{equation}

For each cross-validation iteration, predictions were made on samples not used in training. These predictions were pooled across all folds, and overall R² and RMSE were computed from the aggregated results, ensuring unbiased performance estimation. \\
All analyses were conducted in Python 3.10 using scikit-learn (version 1.3.0) and XGBoost (version 2.0.0).
\section{Results and Discussions}
\subsection{Ensemble Feature Selection}
\begin{sidewaysfigure}
  \centering
  % Top row: Spectral Bands Selection curves 
  \begin{subfigure}[t]{0.2\textheight}
    \centering\includegraphics[width=\linewidth]{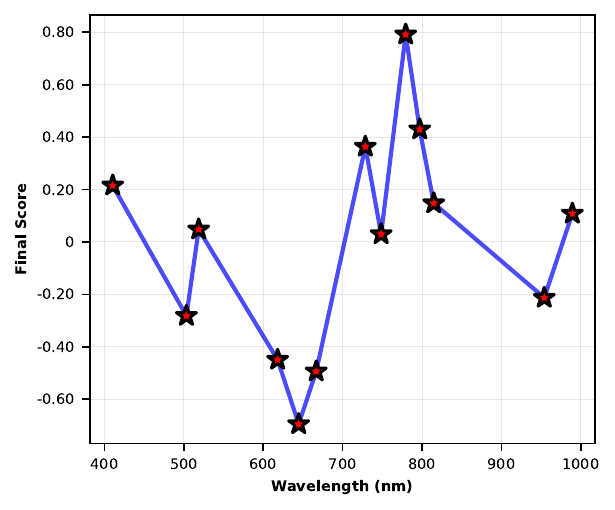}
    \subcaption{Chardonnay\ Leaf Level}
  \end{subfigure}\hspace{0.012\textheight}
  \begin{subfigure}[t]{0.2\textheight}
    \centering\includegraphics[width=\linewidth]{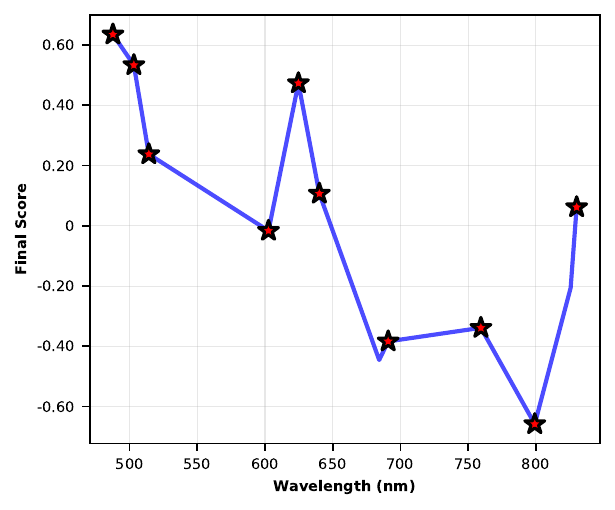}
    \subcaption{Pinot Noir\ Leaf Level}
  \end{subfigure}\hspace{0.012\textheight}
  \begin{subfigure}[t]{0.2\textheight}
    \centering\includegraphics[width=\linewidth]{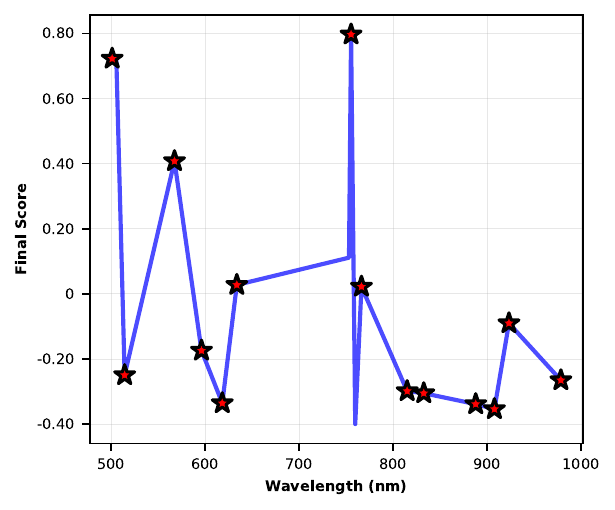}
    \subcaption{Chardonnay\ Canopy Level}
  \end{subfigure}\hspace{0.012\textheight}
  \begin{subfigure}[t]{0.2\textheight}
    \centering\includegraphics[width=\linewidth]{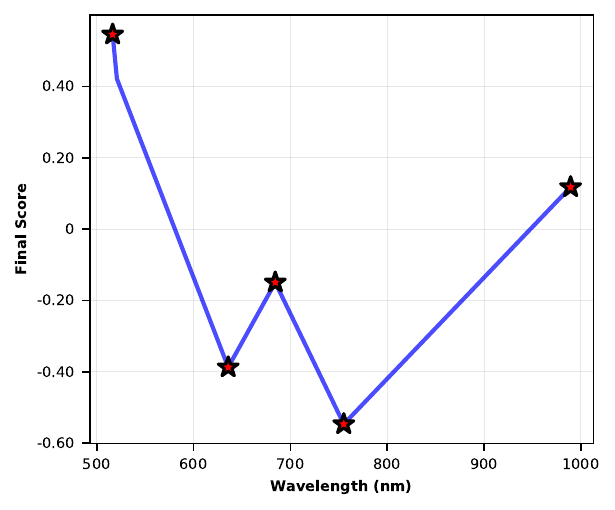}
    \subcaption{Concord\ Canopy Level}
  \end{subfigure}\hspace{0.012\textheight}
  \begin{subfigure}[t]{0.2\textheight}
    \centering\includegraphics[width=\linewidth]{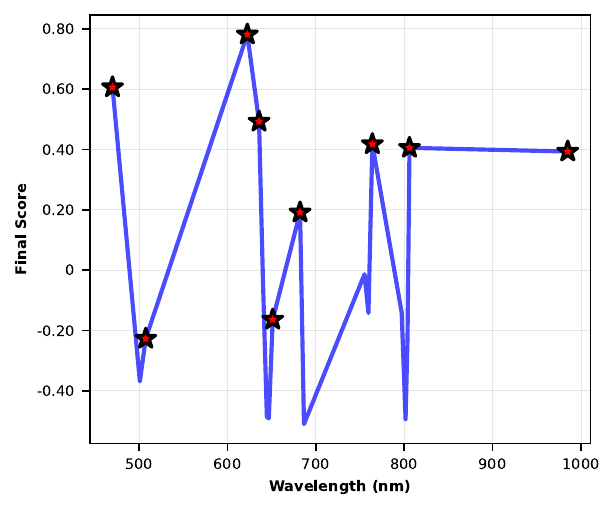}
    \subcaption{Syrah\ Canopy Level}
  \end{subfigure}

  \vspace{0.012\textheight}

  % Bottom row: Stability heatmaps
  \begin{subfigure}[t]{0.2\textheight}
    \centering\includegraphics[width=\linewidth]{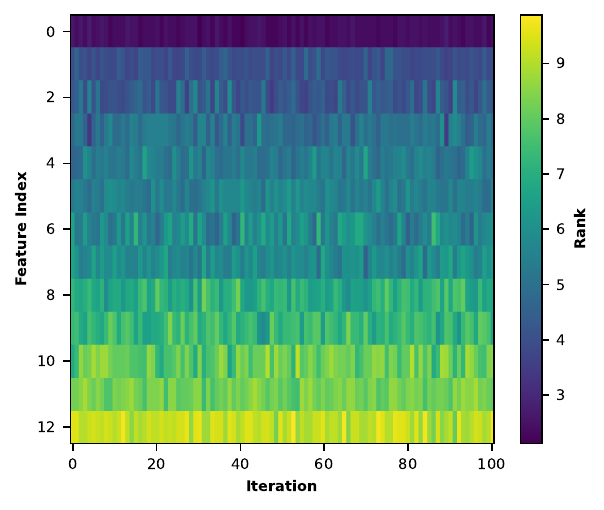}
    \subcaption{Chardonnay\ Leaf Level}
  \end{subfigure}\hspace{0.012\textheight}
  \begin{subfigure}[t]{0.2\textheight}
    \centering\includegraphics[width=\linewidth]{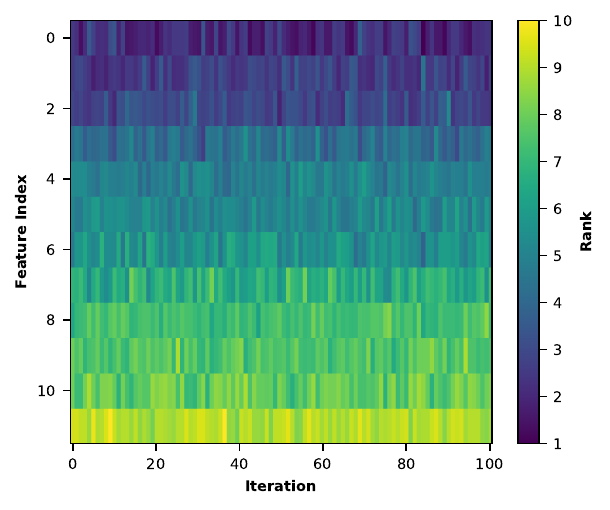}
    \subcaption{Pinot Noir\ Leaf Level}
  \end{subfigure}\hspace{0.012\textheight}
  \begin{subfigure}[t]{0.2\textheight}
    \centering\includegraphics[width=\linewidth]{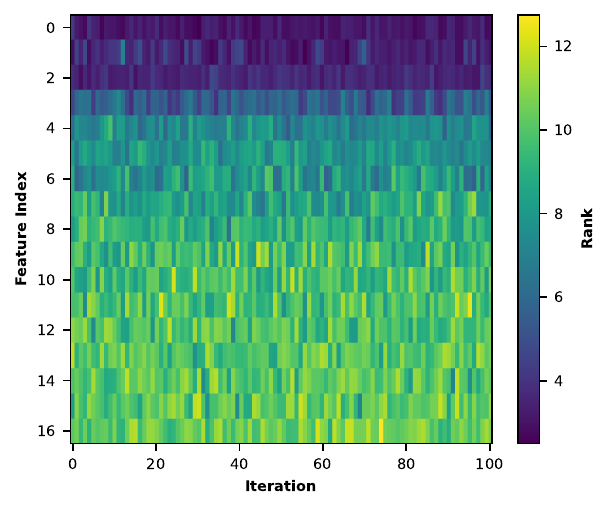}
    \subcaption{Chardonnay\ Ccanopy Level}
  \end{subfigure}\hspace{0.012\textheight}
  \begin{subfigure}[t]{0.2\textheight}
    \centering\includegraphics[width=\linewidth]{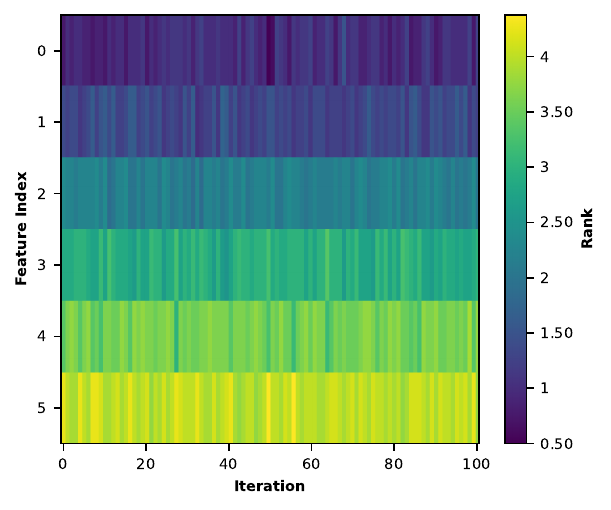}
    \subcaption{Concord\ Canopy Level}
  \end{subfigure}\hspace{0.012\textheight}
  \begin{subfigure}[t]{0.2\textheight}
    \centering\includegraphics[width=\linewidth]{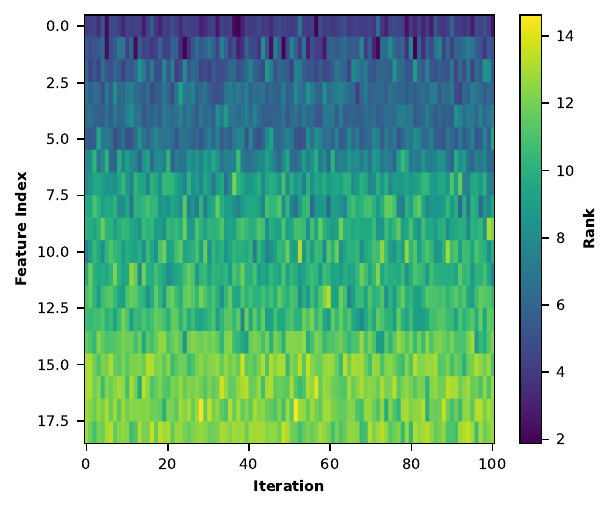}
    \subcaption{Syrah\ Canopy Level}
  \end{subfigure}

  \caption{Final scores and stability heatmaps generated by the comprehensive ensemble feature selection method across all cultivars and two measurement scales (Leaf and Canopy Level). Top (a-e): Ensemble ranking scores. Bottom row (f-j): Ensemble ranking stability heatmaps.}
  \label{fig:all_ensemble}
\end{sidewaysfigure}

The ensemble feature selection method selected different spectral bands for each dataset, demonstrating that both cultivar type and measurement scale are critical factors in identifying the most responsive spectral bands, shown in Figures \ref{fig:all_ensemble}. Across all cultivars, the final ensemble ranking score after the $10nm$ spacing rule is shown in Figure \ref{fig:all_ensemble} (a-e). The heatmaps in Figure \ref{fig:all_ensemble} (f-j) display the selection frequency of cluster representative spectral bands, yielded from the Section \ref{hierarchical_clustering}, across 100 bootstrap iterations, indicating the consistency with which each band maintained its ranking throughout the ensemble feature selection process.

The spectral bands for Chardonnay at canopy level span the entire spectrum, ranging from visible to near-infrared regions. The heatmap shows that the top-ranked spectral bands exhibit consistent stability compared to the lower-ranked ones. In both canopy and leaf level Chardonnay data, the spectral regions show similar patterns, with higher ranking scores in the blue ($450–495nm$) and green ($495–570nm$) regions, and lower scores in the red region ($\approx 590-670nm$); the primary difference occurs in the near infrared region. This difference may be attributed to differences in N ranges across measurement scales. At the leaf level, the lower N range (1.0–2.9\% DW) likely reflects reduced chlorophyll content, potentially increasing the importance of near infrared scattering and yielding higher near infrared scores. In contrast, at the canopy level (2.4–3.7\% DW), higher chlorophyll levels may strengthen absorption in the visible region, possibly shifting feature importance toward visible bands while reducing near infrared contribution \citep{steele2009nondestructive}

At the canopy level, both red cultivars, Concord and Syrah, ensemble rankings showed mixed contributions from visible and red edge bands, with high stability of the top-ranked spectral bands.This pattern could be attributed to anthocyanin--chlorophyll interactions in red grapevine canopies, where the change in N induces pigment shifts that alter reflectance in visible (RGB) more than in white cultivars. The stable selection of broad near infrared features ($750$--$800$$nm$) reflects the contribution of mesophyll scattering and canopy layering, which are positively correlated with N supply \citep{ustin2001simulation}.

\subsection{Optimum Number of Features from Ensemble Feature Selection }
\begin{figure*}[!tbp]
  \centering

  \begin{subfigure}[t]{0.42\textwidth}
    \centering
    \includegraphics[width=\linewidth]{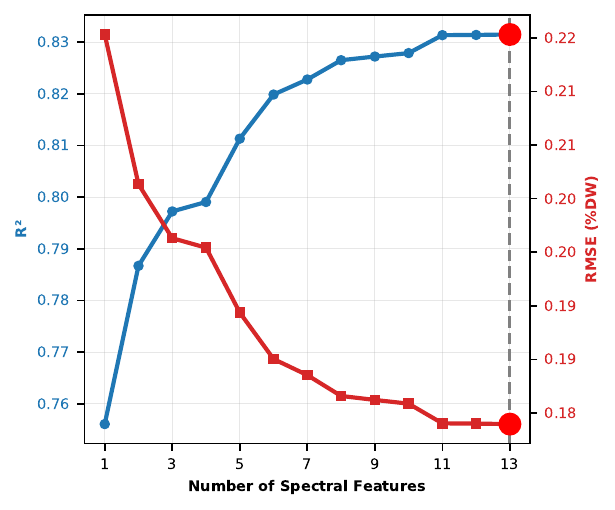}
    \subcaption{Chardonnay Leaf Level}
  \end{subfigure}\hfill
  \begin{subfigure}[t]{0.42\textwidth}
    \centering
    \includegraphics[width=\linewidth]{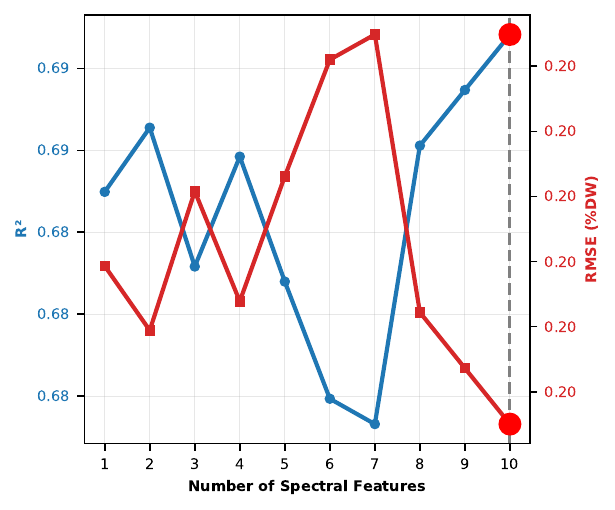}
    \subcaption{Pinot Noir Leaf Level}
  \end{subfigure}

  \vspace{0.4em}

  \begin{subfigure}[t]{0.42\textwidth}
    \centering
    \includegraphics[width=\linewidth]{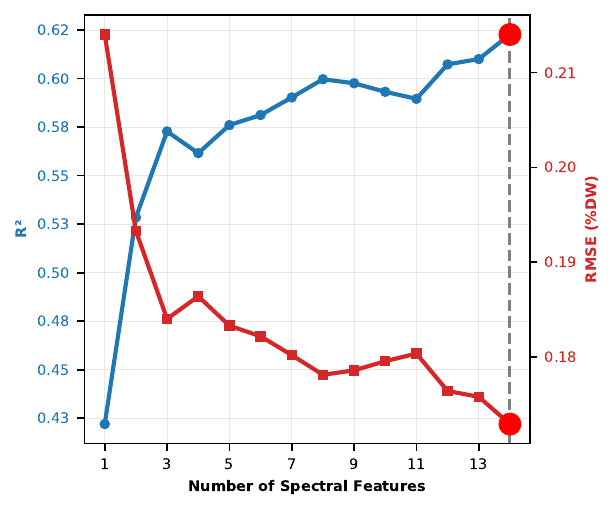}
    \subcaption{Chardonnay Canopy Level}
  \end{subfigure}\hfill
  \begin{subfigure}[t]{0.42\textwidth}
    \centering
    \includegraphics[width=\linewidth]{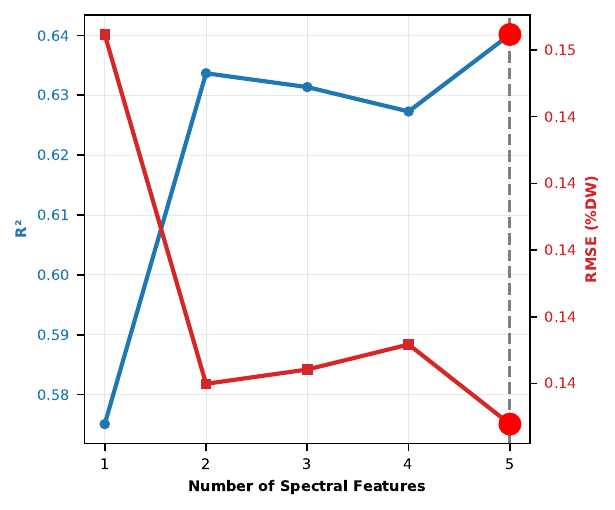}
    \subcaption{Concord Canopy Level}
  \end{subfigure}

  \vspace{0.4em}

  \begin{subfigure}[t]{0.42\textwidth}
    \centering
    \includegraphics[width=0.96\linewidth]{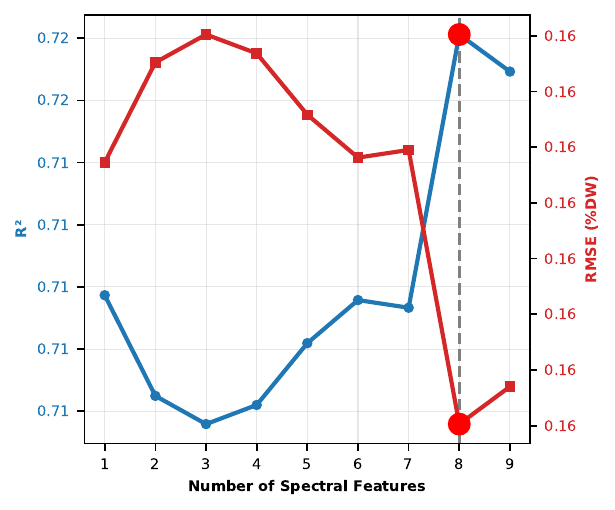}
    \subcaption{Syrah Canopy Level}
  \end{subfigure}

  \caption{Change in N prediction accuracy metrics (RMSE and $R^2$) with increasing number of spectral bands selected by the ensemble feature selection.}
  \label{fig:plsr_opt}
\end{figure*}

The iterative PLSR evaluation with ensemble-ranked spectral bands demonstrated the effectiveness of our feature selection pipeline across cultivars and measurement scales (Fig. \ref{fig:plsr_opt}). In all cases, the first one or two ranked features, combined with the growth stage, already yielded $R^2$ values near $0.5$, indicating that the most responsive spectral bands captured a substantial portion of the N concentration variance. This strong initial performance highlights that the ensemble approach effectively prioritized physiologically meaningful spectral bands, reducing data dimensionality while retaining predictive power.

In the plot, the red line denotes the RMSEs, and the blue line represents R², plotted with an increasing number of features. As additional features were introduced, RMSE decreased and $R^2$ improved, but only up to a point. Beyond 7–10 features, model performance plateaued and in some cases, a slight increase in RMSE was observed. These plots showed that model accuracy generally increased with the number of selected spectral bands, with the highest $R^2$ values reached at the final added feature. 
To evaluate whether additional bands could further improve performance, a benchmark PLSR model was trained using the full spectrum (274 bands) and compared against models using ensemble-selected bands (Table \ref{tab:full_vs_efs}). For Chardonnay leaf level, the 13-band subset outperformed the full spectrum (R² = 0.83 vs. 0.74; RMSE = 0.18\% DW vs. 0.23\%). Similar improvements were observed for Chardonnay canopy level, while Pinot Noir leaf level, Concord canopy level, and Syrah canopy level achieved comparable performance with fewer bands. The improved or equivalent performance of reduced feature sets compared to full-spectrum models suggests that dimensionality reduction minimized noise and redundancy, thereby reducing the risk of overfitting. The final selected spectral bands for each cultivar are listed in Table \ref{tab:selected_spectral_bands}.

\begin{table}[t]
\centering
\caption{PLSR performance with the full spectrum (274 bands) versus ensemble-selected bands. 
Performance measures used are $R^2$ (unitless) and RMSE (\%DW); number of spectral bands of ensemble ranked features used in PLSR for each variety is also reported  (\#bands).}
\label{tab:full_vs_efs}
\begin{tabular}{
l
S[table-format=1.2] S[table-format=1.2]
S[table-format=2.0] S[table-format=1.2] S[table-format=1.2]}
\toprule
& \multicolumn{2}{c}{\textbf{PLSR (274 bands)}} 
& \multicolumn{3}{c}{\textbf{PLSR (Ensemble bands)}} \\
\cmidrule(lr){2-3}\cmidrule(l){4-6}
\multirow{1}{*}{\textbf{Cultivar \& Scale}} 
 & {$R^2$} & {RMSE} 
 & {\#bands} & {$R^2$} & {RMSE} \\
\midrule
Chardonnay leaf level  & 0.74 & 0.23 & 13 & 0.83 & 0.18 \\
Pinot Noir leaf level  & 0.67 & 0.14 & 10 & 0.69 & 0.20 \\
Chardonnay canopy level  & 0.55 & 0.17 & 14 & 0.62 & 0.17 \\
Concord canopy level     & 0.67 & 0.14 &  5 & 0.64 & 0.13 \\
Syrah canopy level       & 0.71 & 0.16 &  8 & 0.72 & 0.16 \\
\bottomrule
\end{tabular}
\end{table}

\begin{table}[H]
\centering
\caption{Final list of ensemble feature selection-ranked spectral bands based on the two performance measures: Maximum $R^2$ and minimum RMSE. The bold spectral bands are top-ranked in each dataset.}
\label{tab:selected_spectral_bands}
\resizebox{\textwidth}{!}{%
\begin{tabular}{|l|p{11.5cm}|c|}
\hline
\textbf{Cultivar \& Scale} & \textbf{Spectral Bands Selected (nm)} & \textbf{Count} \\
\hline
Chardonnay leaf level  & 410.28, 503.11, 518.59, 618.05, 644.58, 666.68, 728.57, 748.46, $\textbf{779.41}$, 797.09, 814.78, 954.03, 989.39 &  13 \\
Pinot Noir leaf level  & $\textbf{487.64}$, 503.11, 514.17, 602.58, 624.68, 640.16, 690.99, 759.52, 799.3, 830.25 & 10 \\
Chardonnay canopy level  & 500.9, 514.17, 567.22, 595.95, 618.05, 633.53, $\textbf{755.09}$, 766.15, 814.78, 832.46, 887.72, 907.61, 923.08, 978.34 &  14 \\
Concord canopy level     & $\textbf{516.38}$, 635.74, 684.36, 755.09, 989.39 & 5 \\
Syrah canopy level       & 469.96, $\textbf{622.47}$, 635.74, 651.21, 682.15, 763.94, 805.93, 984.97 & 8 \\
\hline
\end{tabular}%
}
\end{table}

\subsection{Nitrogen Prediction}
\subsubsection{Leaf Level Analysis}
\begin{figure}[!tbp]
  \centering
  % Row 1
  \begin{subfigure}[t]{0.48\textwidth}
    \centering
    \includegraphics[width=\linewidth]{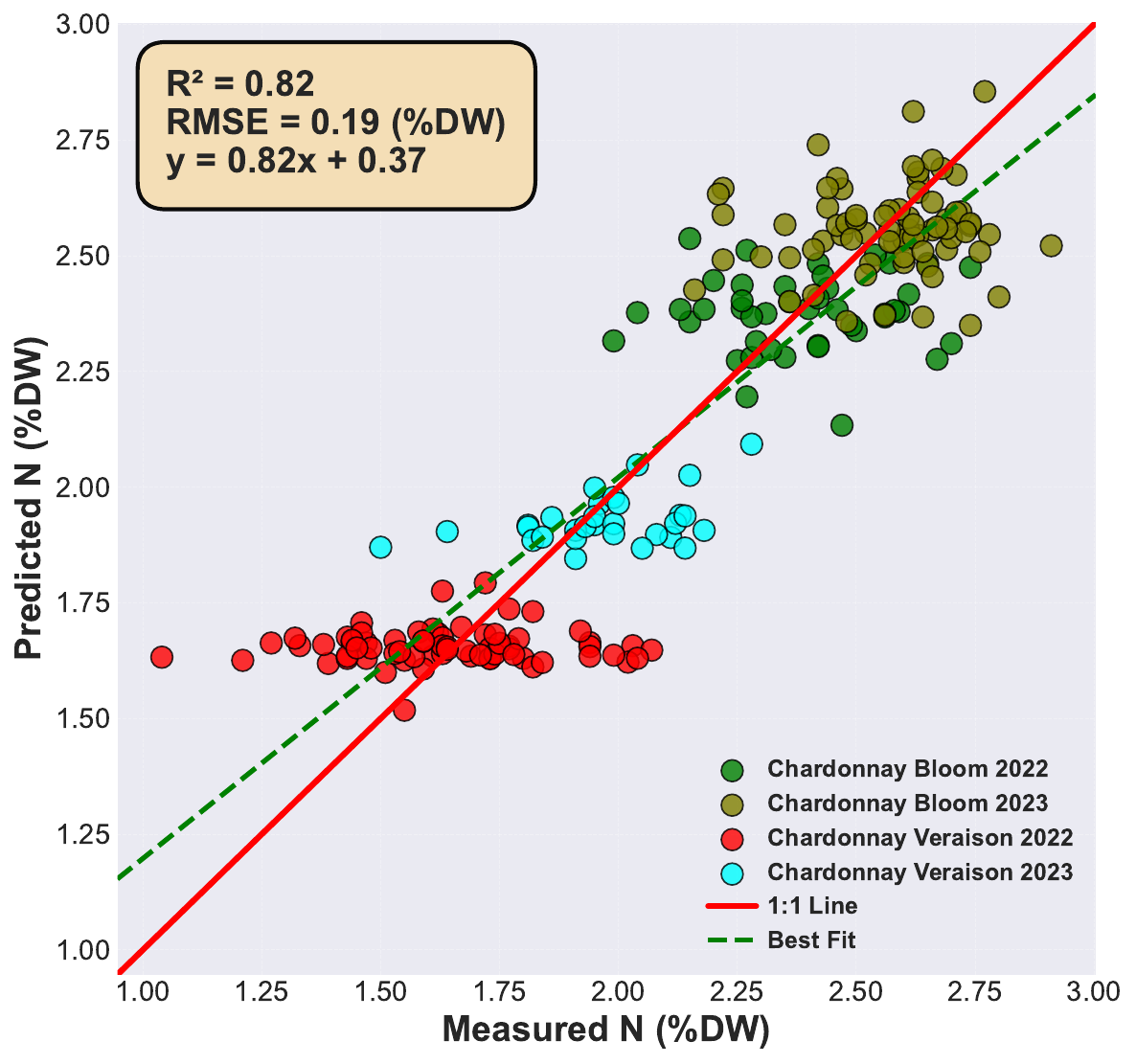}
    \subcaption{SVR-RBF (Chardonnay leaf level)}
    \label{fig:CH_LL_GPR}
  \end{subfigure}\hfill
  \begin{subfigure}[t]{0.48\textwidth}
    \centering
    \includegraphics[width=\linewidth]{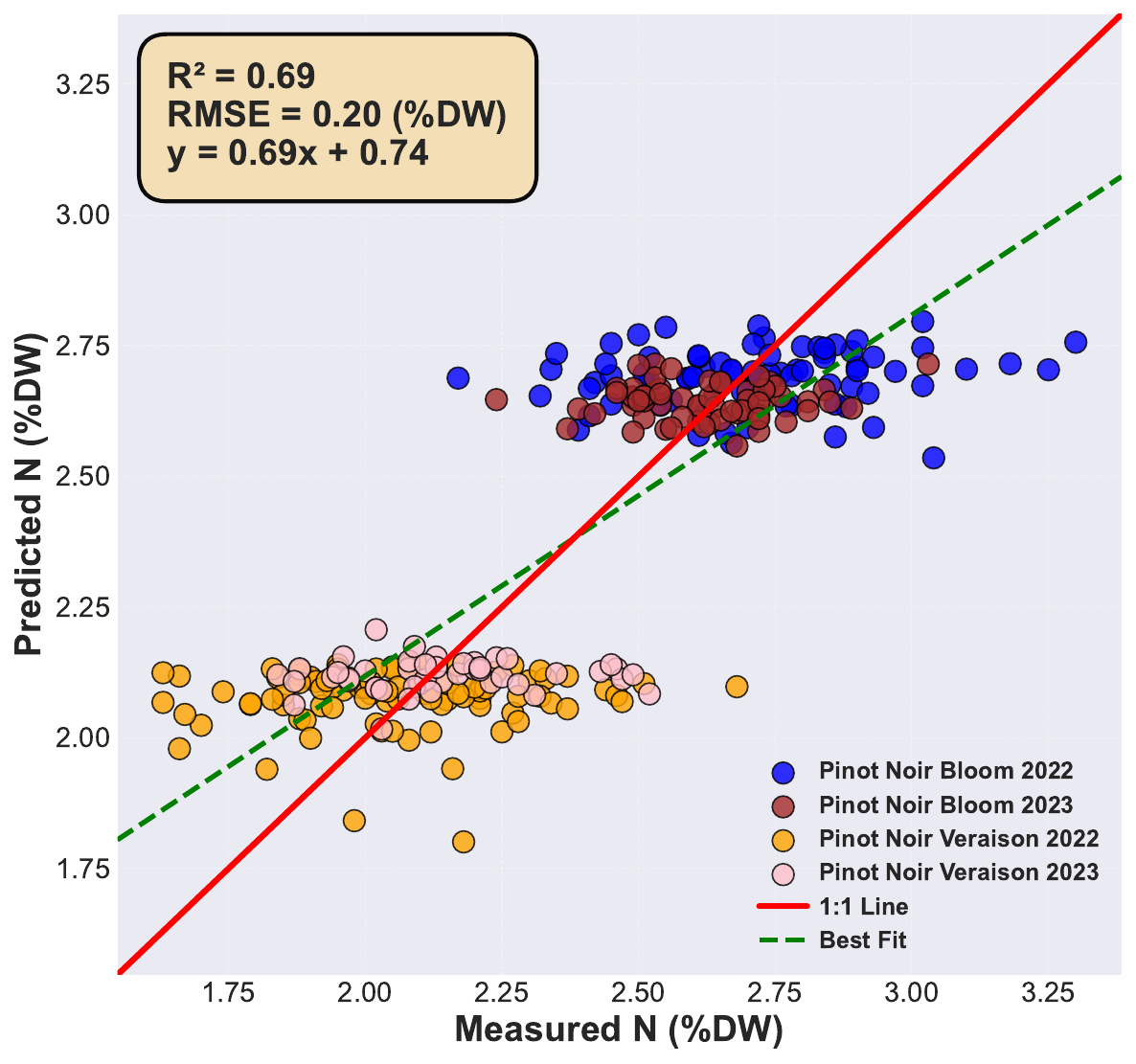}
    \subcaption{Gaussian Process Regressor (Pinot Noir leaf level)}
    \label{fig:PN_LL_GPR}
  \end{subfigure}
  \caption{Leaf-level N concentration estimation for two cultivars:
  (a) SVR-RBF for Chardonnay, (b) Gaussian Process Regressor for Pinot Noir. Each plot showed the predicted vs.\ measured N (\%DW); the green line is the best-fit regression and the red line is the 1:1 reference.}
  \label{fig: LL_N_models}
\end{figure}

The performance of N concentration prediction models at the leaf level using ensemble-selected spectral bands is shown in Figure \ref{fig: LL_N_models}. For Chardonnay, the SVR-RBF regression achieved better accuracy ($R^2$ = 0.82, RMSE = 0.19\% DW), closely aligning predicted and measured values across both seasons and growth stages. For Pinot Noir, Gaussian Process Regression (GPR) yielded ($R^2$ = 0.69, RMSE = 0.20\% DW), with slightly larger residual variation compared to Chardonnay. Performance metrics for all ML models for Chardonnay and Pinot Noir at leaf level are summarized in Table \ref{tab:leaf_models_en_xgb_svr_gpr}.
This result showed a noticeable improvement in leaf level predictions compared to those at the canopy level. This enhanced predictive performance at the leaf level can be attributed to several factors. The wider range of N concentrations in leaf level data provided more comprehensive training data for the models. Furthermore, leaf level spectral data significantly reduced the effects of variable leaf orientation and inconsistent sunlight exposure that typically complicate canopy level spectral data. The more consistent light incidence at the leaf level produced spectral signatures with reduced environmental noise, which helped improve model performance in estimating leaf N concentration, highlighting the potential of leaf level spectral analysis for precise N status assessment.

\begin{table}[t]
\centering
\caption{Model performance ($R^2$ and RMSE \%DW) for Leaf-level N estimation in two grapevine cultivars (Chardonnay, Pinot Noir) using Elastic Net, XGBoost, SVR–RBF, and Gaussian Process Regression (GPR) models. Each model was trained with ensemble-selected spectral bands for an individual cultivar.}
\label{tab:leaf_models_en_xgb_svr_gpr}
\begin{tabular}{lcccccccc}
\toprule
\multirow{2}{*}{Cultivar (Leaf level)} 
& \multicolumn{2}{c}{\textbf{Elastic Net}} 
& \multicolumn{2}{c}{\textbf{XGBoost}} 
& \multicolumn{2}{c}{\textbf{SVR–RBF}} 
& \multicolumn{2}{c}{\textbf{GPR}} \\
\cmidrule(lr){2-3}\cmidrule(lr){4-5}\cmidrule(lr){6-7}\cmidrule(l){8-9}
& {$R^2$} & {RMSE} & {$R^2$} & {RMSE} & {$R^2$} & {RMSE} & {$R^2$} & {RMSE} \\
\midrule
Chardonnay & {0.79} & {0.19} & {0.78} & {0.20} & {0.82} & {0.19} & {0.81} & {0.19} \\
Pinot Noir & {0.68} & {0.19} & {0.66} & {0.20} & {0.68} & {0.19} & {0.69} & {0.19} \\
\bottomrule
\end{tabular}
\end{table}

\subsubsection{Canopy Level Analysis}
The predictive performance of canopy-level N estimation models using selected features is summarized in Table \ref{tab: CL_N_all_models}. Among the four ML models evaluated, GPR performed best for Chardonnay and Concord, achieving $R^2$ values of 0.65 (RMSE = 0.17\%DW) for Chardonnay and 0.72 (RMSE = 0.12\%DW) for Concord whereas Elastic Netperformed best for Syrah among other models with an $R^2$of 0.70 (RMSE = 0.16\%DW). Performance differences across cultivars might have caused both by the measured N range and the physiological characteristics of each cultivar. Concord and Syrah, with higher mean N concentration values ($\approx$3.0–3.12 \%DW), achieved greater predictive accuracy compared to Chardonnay ($\approx$2.91 \%DW). In Chardonnay, the N distribution was more skewed to one end, introducing additional variability in the reflectance-N relationship and reducing overall model fit. Accuracy differences also reflect the influence of selected spectral features. Chardonnay models relied more heavily on near infrared bands, which are inherently more sensitive to external factors such as canopy structure, leaf angle, and water status. In contrast, in Concord and Syrah (both red cultivars), rely more on visible bands, which provided a more stable signal for N prediction. An additional complication arises from nutrient interactions: spectral similarities caused by correlated deficiencies can mask N-specific responses, as many deficiencies reduce chlorophyll concentration and produce overlapping signatures \citep{rustioni2018iron, chancia2021assessing}.

\begin{table}[t]
\centering
\caption{Model performance ($R^2$ and RMSE \%DW) for canopy-level N estimation in three grapevine cultivars (Chardonnay, Concord, and Syrah) using Elastic Net, XGBoost, SVR–RBF, and Gaussian Process Regression (GPR). Each model was trained with ensemble-selected spectral bands for an individual cultivar.}
\label{tab: CL_N_all_models}
\begin{tabular}{lcccccccc}
\toprule
\multirow{2}{*}{Cultivar} 
& \multicolumn{2}{c}{\textbf{Elastic Net}} 
& \multicolumn{2}{c}{\textbf{XGBoost}} 
& \multicolumn{2}{c}{\textbf{SVR–RBF}} 
& \multicolumn{2}{c}{\textbf{GPR}} \\
\cmidrule(lr){2-3}\cmidrule(lr){4-5}\cmidrule(lr){6-7}\cmidrule(l){8-9}
& {$R^2$} & {RMSE} & {$R^2$} & {RMSE} & {$R^2$} & {RMSE} & {$R^2$} & {RMSE} \\
\midrule
Chardonnay & {0.60} & {0.18} & {0.64} & {0.17} & {0.63} & {0.17} & {0.65} & {0.17} \\
Concord    & {0.66}   & {0.13}   & {0.71}   & {0.12}   & {0.72}   & {0.12}   & {0.72} & {0.12} \\
Syrah      & {0.70}   & {0.16}   & {0.68}   & {0.17}   & {0.65}   & {0.18}   & {0.68} & {0.17} \\
\bottomrule
\end{tabular}
\end{table}

\begin{figure}[!tbp]
  \centering

  % Row 1
  \begin{subfigure}[t]{0.48\textwidth}
    \centering
    % <-- replace with your actual filename
    \includegraphics[width=\linewidth]{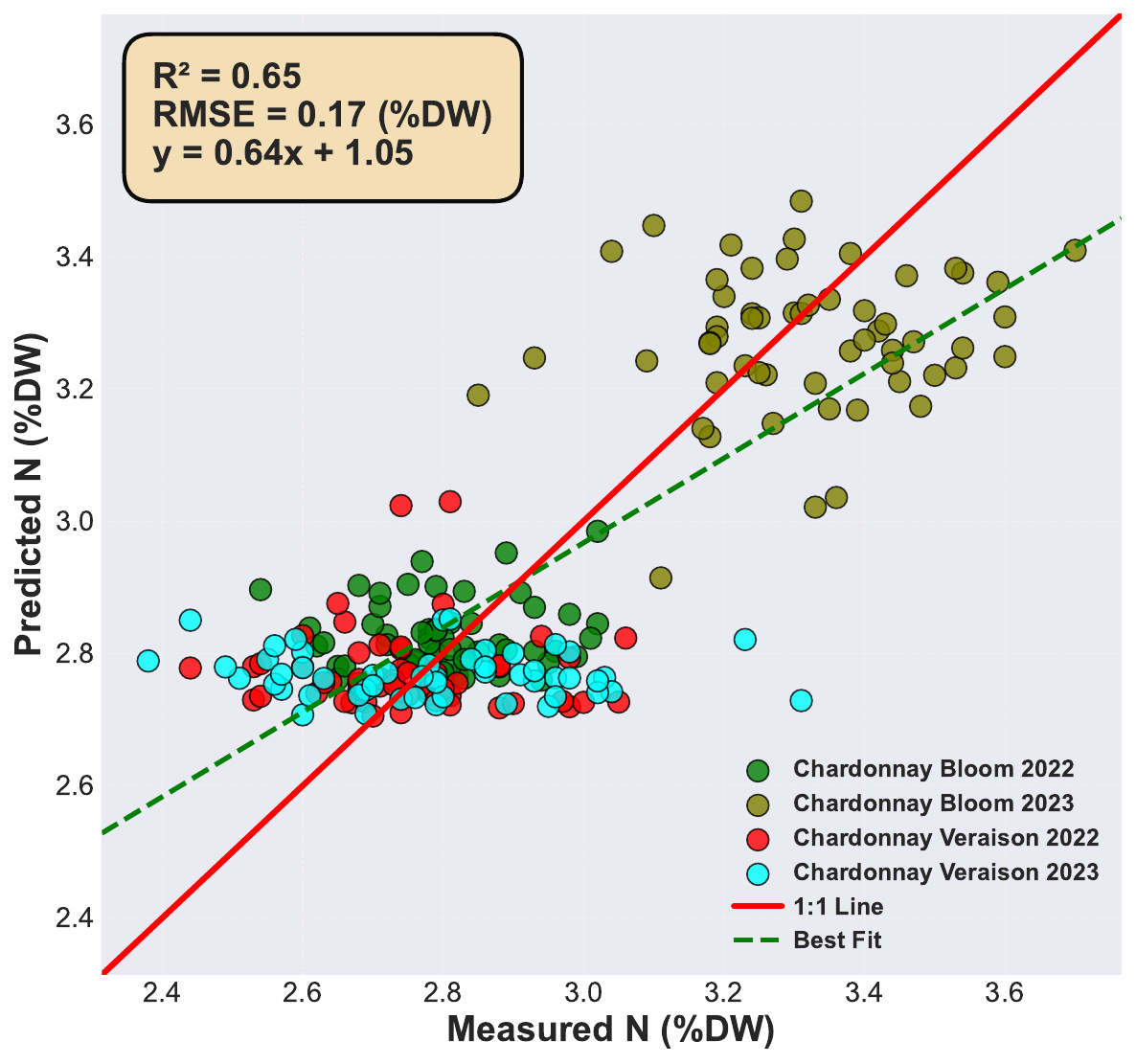}
    \subcaption{Gaussian Process Regressor (Chardonnay Canopy Level)}
    \label{fig:CH_CL_GPR}
  \end{subfigure}\hfill
  \begin{subfigure}[t]{0.48\textwidth}
    \centering
    % <-- replace with your actual filename
    \includegraphics[width=\linewidth]{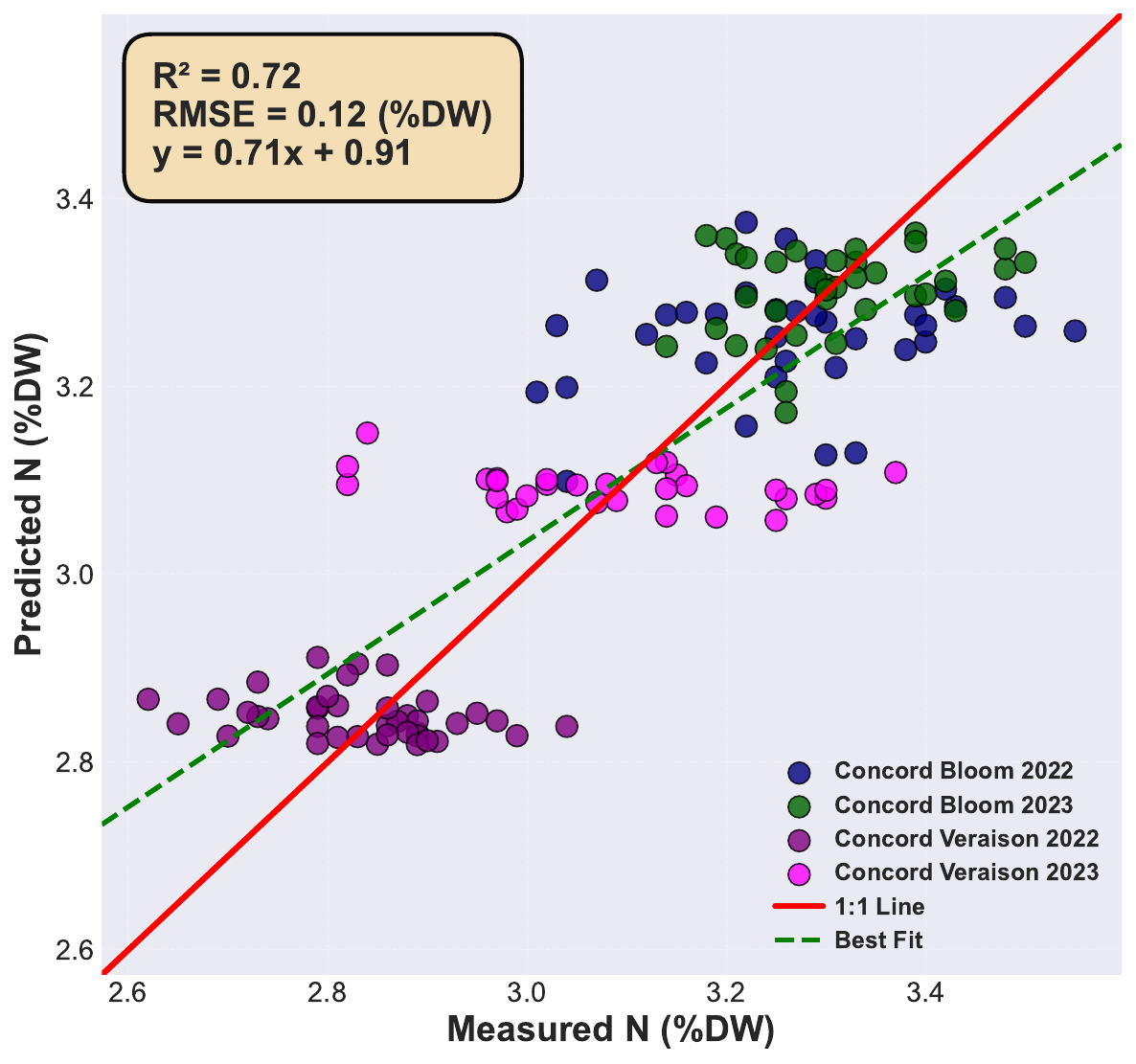}
    \subcaption{Gaussian Process Regressor (Concord Canopy Level)}
    \label{fig:CON_LL_GPR}
  \end{subfigure}

  \vspace{0.4em}

  % Row 2 (single wide to avoid an empty cell)
  \begin{subfigure}[t]{0.48\textwidth}
    \centering
    % <-- replace with your actual filename
    \includegraphics[width=0.98\linewidth]{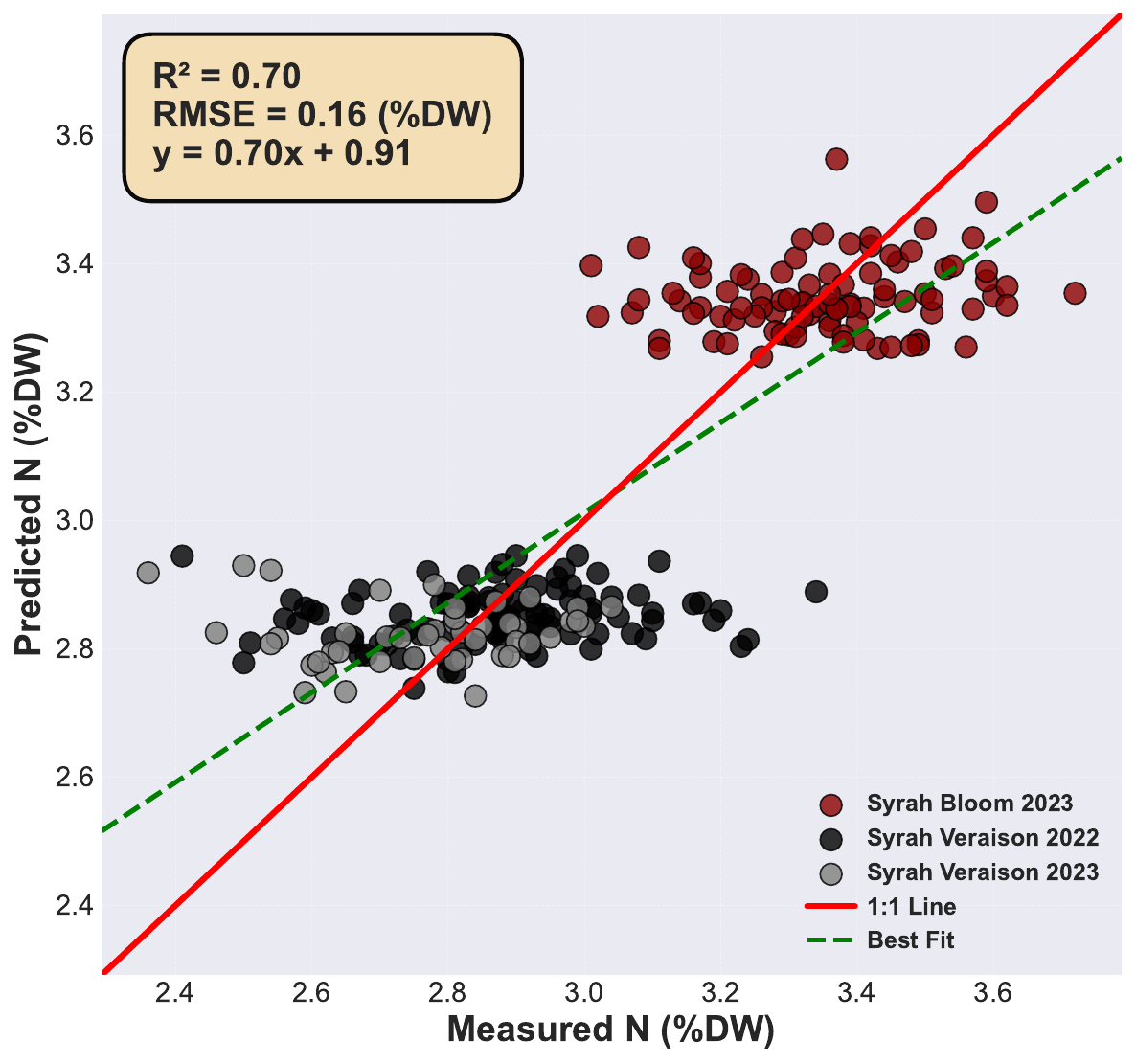}
    \subcaption{ElasticNet (Syrah Canopy Level)}
    \label{fig:SY_LL_GPR}
  \end{subfigure}

  \caption{Canopy-level N estimation for three cultivars: (a) Gaussian Process Regressor for Chardonnay, (b) Gaussian Process Regressor for Concord, and (c) Elastic Net for Syrah. Each plot shows predicted vs.\ measured N (\%DW); the green line is the best-fit regression and the red line is the 1:1 reference.}\label{fig:nitrogen_histo}
  \label{fig: CL_N_models}
\end{figure}

Additionally, environmental variables such as lighting conditions and leaf orientation introduce variability in spectral data collection \citep{kang2023estimating, kang2024assessing}. Despite these challenges, these models validate ground-based hyperspectral imaging as an effective method for non-destructive N assessment in grapevine leaves, demonstrating robust performance across multiple cultivars, and growing seasons.
\subsubsection{Leaf-to-Canopy Band Transferability}
Hyperspectral imaging at the leaf level provided better spectral–physiological relationships due to consistent illumination and leaf angle, and reduced overall external noise. However, applying these relationships at the canopy scale introduces additional complexity due to variable leaf angles, shadowing, canopy layering, and often background reflectance. The transferability and robustness of the feature selection approach were evaluated by testing whether the N-sensitive bands identified at the leaf level remained predictive at the canopy level. To further isolate physiological effects from external noise, we focus on the transferability analysis on one white cultivar (Chardonnay) and one red cultivar (Pinot Noir), ensuring that any observed change in predictive performance primarily reflects physiological rather than environmental or measurement scale factors. These cultivars exhibit contrasting leaf pigment profiles—Chardonnay, lacking anthocyanins, and Pinot Noir, rich in them, allowing us to assess whether the ensemble-selected N-sensitive bands are robust to pigment-related spectral variability. In particular, several wavelength regions identified at the leaf level were also retained at the canopy level, demonstrating partial consistency in N-sensitive spectral responses across scales. 

\begin{table}[t]
\centering
\caption{Model performance ($R^2$ and RMSE, \%DW) for canopy-level N estimation in three grapevine cultivars (Chardonnay, Concord, and Syrah) using Elastic Net, XGBoost, SVR–RBF, and Gaussian Process Regression (GPR). Each model was trained with leaf-level spectral bands selected from Chardonnay (white) and Pinot Noir (red) to test the cross-scale and cross-cultivar transferability of the selected features from leaf to canopy.}
\label{tab: CL_N_LL_bands}
\begin{tabular}{lcccccccc}
\toprule
\multirow{2}{*}{Cultivar} 
& \multicolumn{2}{c}{\textbf{Elastic Net}} 
& \multicolumn{2}{c}{\textbf{XGBoost}} 
& \multicolumn{2}{c}{\textbf{SVR–RBF}} 
& \multicolumn{2}{c}{\textbf{GPR}} \\
\cmidrule(lr){2-3}\cmidrule(lr){4-5}\cmidrule(lr){6-7}\cmidrule(l){8-9}
& {$R^2$} & {RMSE} & {$R^2$} & {RMSE} & {$R^2$} & {RMSE} & {$R^2$} & {RMSE} \\
\midrule
Chardonnay & {0.67} & {0.16} & {0.64} & {0.17} & {0.65} & {0.17} & {0.65} & {0.17} \\
Concord    & {0.65}   & {0.13}   & {0.70}   & {0.12}   & {0.68}   & {0.13}   & {0.71} & {0.12} \\
Syrah      & {0.69}   & {0.17}   & {0.68}   & {0.17}   & {0.69}   & {0.17}   & {0.70} & {0.16} \\
\bottomrule
\end{tabular}
\end{table}

This focused comparison provides a meaningful test of both within-cultivar cross-scale transferability (Chardonnay leaf level → Chardonnay canopy level) and across-cultivar, same-pigmentation transferability (Pinot Noir leaf level → Syrah and Concord canopy level). The canopy level N prediction using models trained on canopy level selected spectral bands (Table \ref{tab: CL_N_all_models}), the canopy level N prediction using models trained on leaf level selected spectral bands showed comparablemade accuracy, indicating good cross-scale transferability of the selected features. For instance, canopy level N in Chardonnay was predicted with an $ R^2$ of 0.67 using leaf level selected bands, compared to 0.65 with canopy level selected bands, indicating a slightly better predictive strength. Concord and Syrah also retained comparable accuracy ($R^2 \approx 0.70$--$0.71$), only slightly below or similar to models using their respective canopy level selected bands. These results suggest that many physiologically relevant spectral bands identified at the leaf level—particularly those in the visible ($500$--$670\,\mathrm{nm}$), red edge ($700$--$760\,\mathrm{nm}$) and near infrared ($\approx 800$--$830\,\mathrm{nm}$, $\approx 950$--$990\,\mathrm{nm}$) regions remained predictive for N when scaled up to canopy reflectance. The strong performance across cultivars demonstrated that the ensemble-selected spectral bands captured robust, N-sensitive spectral traits rather than scale or cultivar-specific artifacts, supporting their potential for transferability.
This finding addresses a critical gap in vineyard hyperspectral imaging research. Previous hyperspectral studies for grapevine N assessment have employed UAS-based (aerial) platforms, with limited investigation of ground-based hyperspectral imaging and cross-scale transferability \citep{peanusaha2024nitrogen, chancia2021assessing, pourreza2025nitrogen}. \citet{peanusaha2024nitrogen} achieved $R^2$ = 0.78 for leaf-level N prediction using handheld spectroradiometry across two table grape cultivars, while \citet{pourreza2025nitrogen} reported $R^2$ = 0.68–0.69 for canopy-level N assessment using UAS-based hyperspectral imaging on a single table grape cultivar. \citet{chancia2021assessing} employed ensemble feature selection with UAS hyperspectral imagery (VNIR + SWIR) to identify N-sensitive bands, achieving $R^2$ = 0.44–0.45 at canopy level across Concord vineyards. While these UAS-based approaches demonstrate the potential of hyperspectral remote sensing for N monitoring, ground-based hyperspectral imaging remains underexplored for vineyard N assessment. Furthermore, systematic evaluation of whether N-sensitive spectral features identified at leaf level retain predictive capability when applied to canopy-scale measurements has received limited attention. The present study addresses both gaps, demonstrating that ground-based hyperspectral imaging effectively assesses N across measurement scales and that leaf-level selected spectral bands achieve $R^2$ = 0.64–0.71 for canopy-level N prediction using raw reflectance values.

\begin{figure}[!tbp]
  \centering

  % Row 1
  \begin{subfigure}[t]{0.48\textwidth}
    \centering
    % <-- replace with your actual filename
    \includegraphics[width=\linewidth]{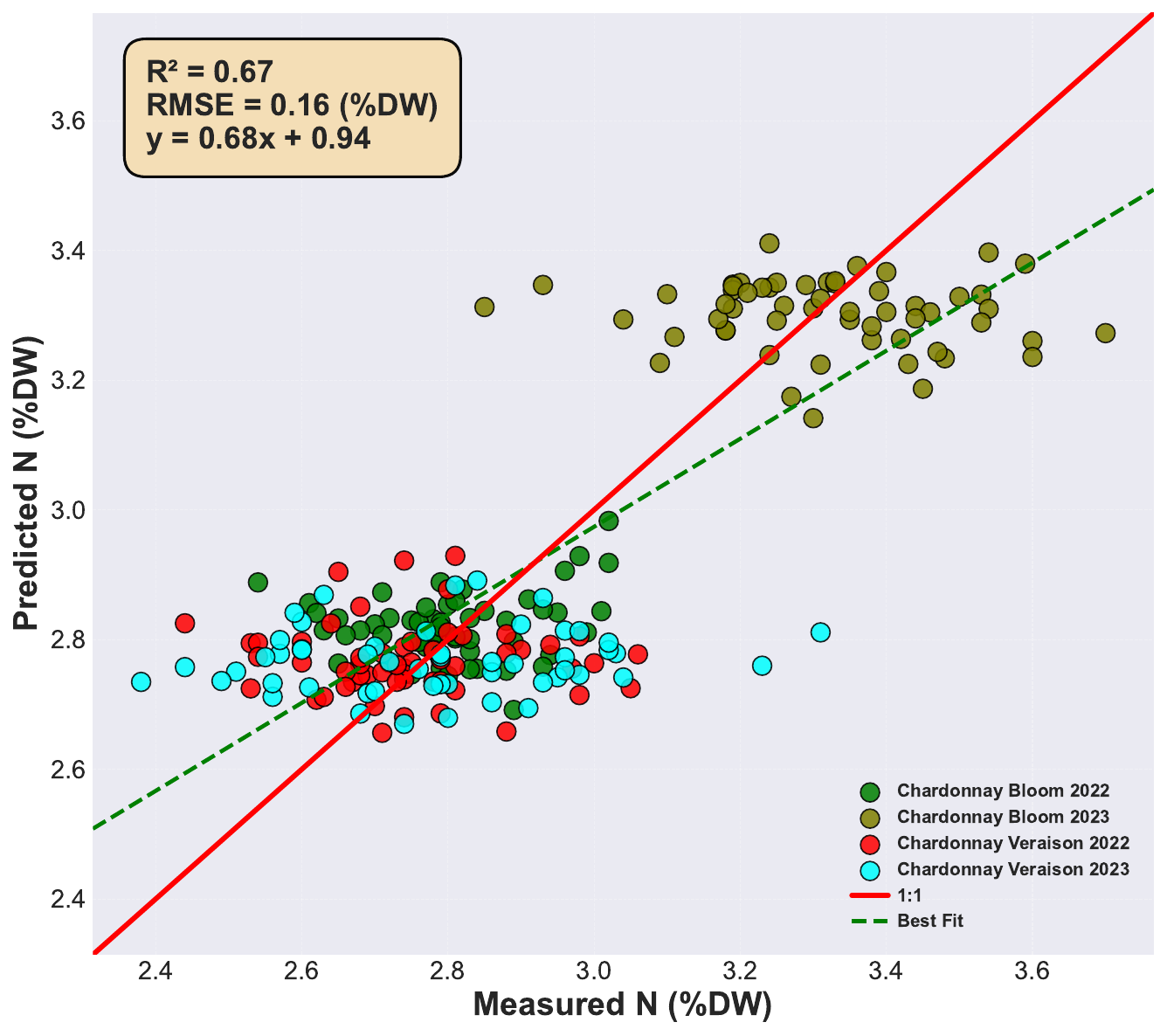}
    \subcaption{ElasticNet (Chardonnay Canopy Level)}
    \label{fig:CH_CL_ElasticNet_with_LL_bands}
  \end{subfigure}\hfill
  \begin{subfigure}[t]{0.48\textwidth}
    \centering
    % <-- replace with your actual filename
    \includegraphics[width=\linewidth]{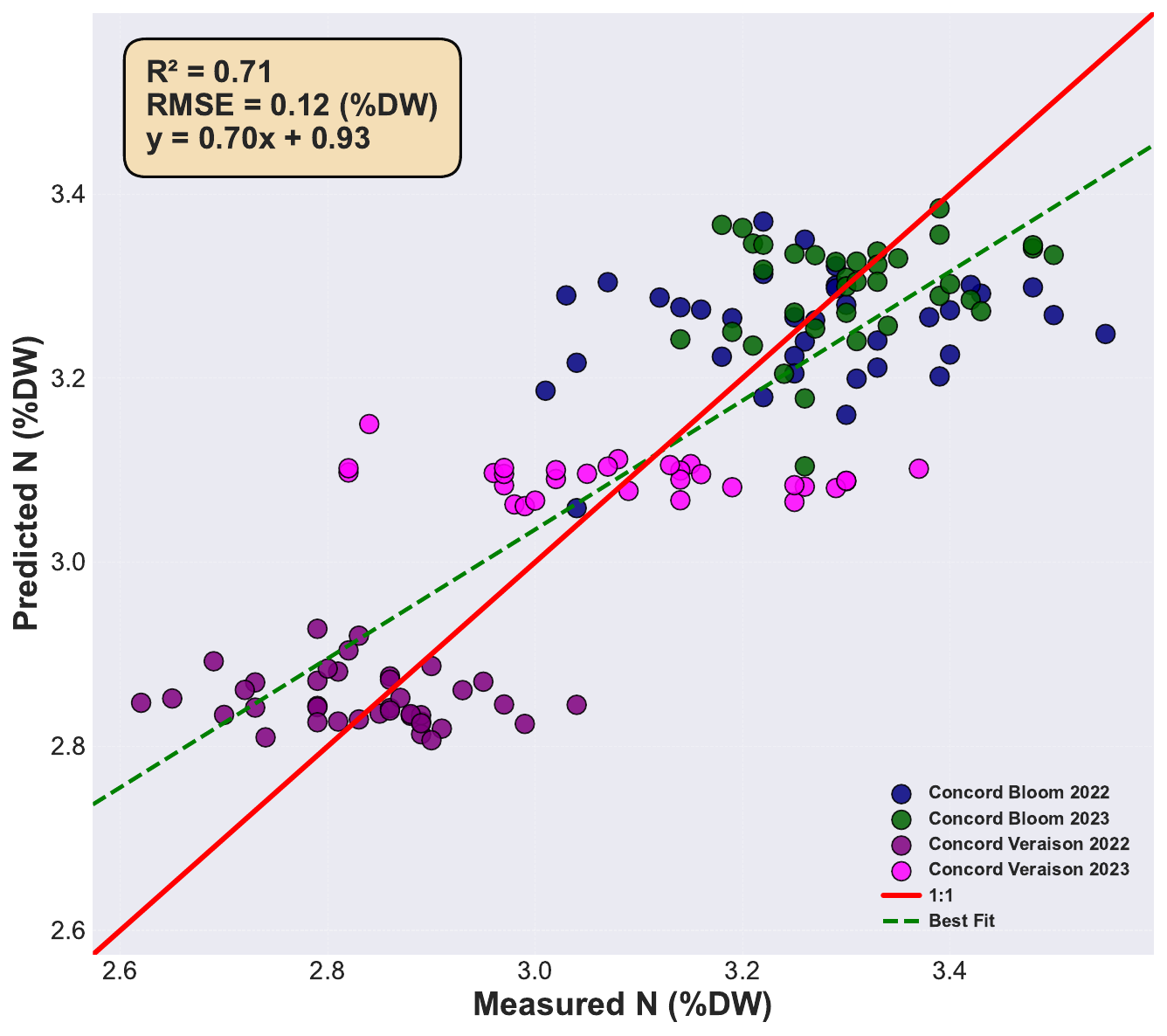}
    \subcaption{Gaussian Process Regressor (Concord Canopy Level)}
    \label{fig:CON_LL_GPR_with_LL_bands}
  \end{subfigure}

  \vspace{0.4em}

  % Row 2 (single wide to avoid an empty cell)
  \begin{subfigure}[t]{0.48\textwidth}
    \centering
    % <-- replace with your actual filename
    \includegraphics[width=0.98\linewidth]{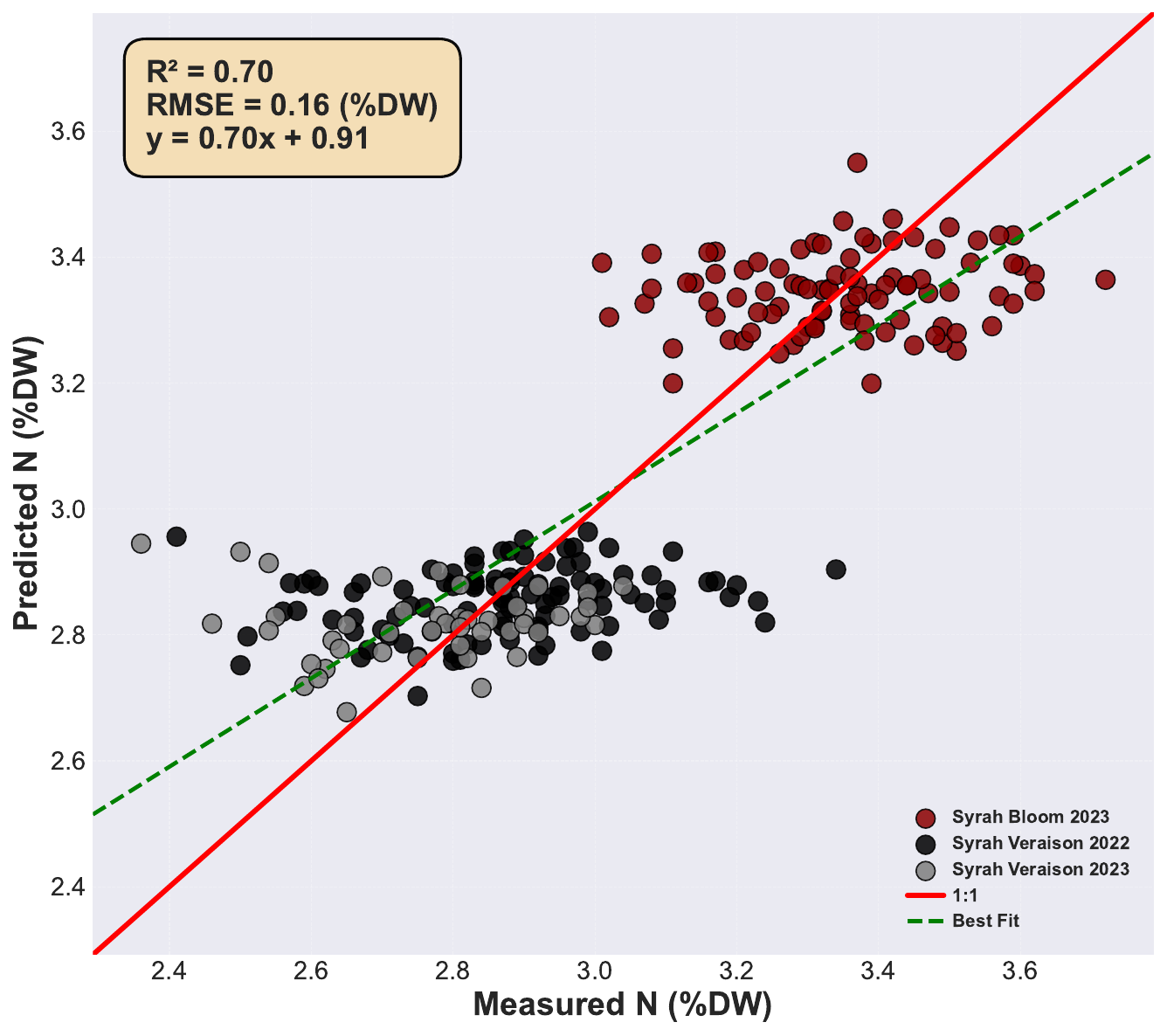}
    \subcaption{Gaussian Process Regressor (Syrah Canopy Level)}
    \label{fig:SY_LL_GPR_with_LL_bands}
  \end{subfigure}

  \caption{Canopy-level N estimation for three cultivars: (a) Gaussian Process Regressor for Chardonnay, (b) Gaussian Process Regressor for Concord, and (c) Elastic Net for Syrah. Each plot shows predicted vs.\ measured N (\%DW); the green line is the best-fit regression and the red line is the 1:1 reference.}\label{fig:nitrogen_histo2}
  \label{fig: CL_N_models_2}
\end{figure}

Beyond cross-scale transferability, the models successfully captured biological patterns in seasonal N dynamics. Nitrogen concentrations in grapevine leaves measured using laboratory chemical analysis were consistently higher during the bloom stage compared to the veraison stage. This pattern was consistent in both leaf and canopy level analyses across all grapevine cultivars tested (Chardonnay, Concord, Syrah, Pinot Noir), with the exception of Chardonnay in 2022, as explained in Section \ref{TN_CL} and illustrated in Figure \ref{fig:TN_CL_histo}. This seasonal N decline aligns with findings reported in previous studies \citep{verdenal2021understanding, keller2020science}. Importantly, our prediction models successfully captured this biological trend across different datasets. This consistent pattern across multiple cultivars (grown in different vineyards) and growing seasons suggests that combining data from both bloom and veraison stages strengthens model robustness rather than developing growth-stage-specific models. While bloom-stage measurements generally showed higher N concentrations, incorporating both stages in our predictive models provided several advantages: (1) it captured the full range of N variability throughout the growing season, (2) it increased sample size and diversity for more robust model training, and (3) it enabled the model to account for seasonal N dynamics when making predictions. Furthermore, since grapevines are significantly affected by year-to-year weather variations, incorporating data from multiple growing seasons (2022 and 2023) enhanced the model generalizability across different environmental conditions. This multi-cultivar, multi-stage, and multi-season approach led to more robust models capable of maintaining prediction accuracy despite the inherent variability in vineyard systems due to soil and climate fluctuations, making them more reliable tools for practical N management across diverse growing conditions and years. \\
By incorporating three wine and juice grape cultivars grown in different climatic regions (i.e., cool-humid western Oregon versus warm-arid eastern Washington) over two growing seasons and two phenological stages, this study extended previous research on single cultivars (e.g., \citet{chancia2021assessing, pourreza2025nitrogen}). This multi-cultivar, multi-environment validation enhanced the robustness and cross-cultivar transferability of our results. The implementation of rigorous cross-validation ensures the models' generalizability across diverse vineyard conditions, enhancing their practical applicability. This study provided a rapid, non-destructive assessment of N for grapevine leaves, which has a potential to substantially reduce labor in sampling effort compared to conventional methods. Further improvements in prediction accuracy could be achieved by incorporating complementary data, such as soil properties, weather conditions, and vine phenology, and by combining sensors with other remote sensing technologies.

\section{Conclusion}
This research work developed a practical, noninvasive method for estimating N in grapevine leaves. An ensemble feature selection pipeline was developed, which yielded compact, physiologically meaningful bands. Following the feature selection, lightweight ML models were trained, which performed well in predicting vine N levels at both leaf and canopy scales. Across cultivars, the selected spectral regions with responsiveness to N were: visible ($\approx500-670nm$; chlorophyll absorption), red edge ($\approx700-760nm$; chlorophyll/red edge position), and near infrared ($> 760nm$; leaf internal structure/water). A clear cultivar pattern emerged: white grapes (Chardonnay) showed an almost balanced proportion of selected spectral bands across visible, red edge, and near infrared regions, whereas the spectral bands selected for red grapes (Pinot Noir, Syrah, Concord) were concentrated more in the visible, reflecting anthocyanin-chlorophyll interactions that dominated their visible response. Between scales, leaf level models used more bands and achieved better model performance for estimating N than canopy level, which was primarily caused by leaf spectra preserving fine pigment/leaf structure cues; consequently, canopy level selection favors fewer, more robust visible/red edge features. The top ranked bands for each cultivar (e.g., $779.41nm$ for Chardonnay leaf level, $487.64nm$ for Pinot Noir leaf level, $755.09nm$ for Chardonnay canopy level, $516.38/622.47nm$ for Concord/Syrah canopy level) are the top-ranked spectral bands in ensemble results; notably, combining growth stage with these top ranked bands already produced $\sim0.5$ $R^2$, because (i) growth stage explains phenology-driven baseline shifts in N/chlorophyll, and (ii) a single, physiology-anchored band (typically a red edge for white cultivar or strong visible absorption for red cultivars) captures a large share of the remaining variance. Collectively, these results demonstrate that a compact and transferable set of spectral bands, balanced across the visible, red edge, and near infrared regions for white cultivars and more focused on the visible region for red cultivars, can reliably estimate N in grapevine leaves. Based on these results, we suggest future research investigate the mechanistic basis for cultivar-specific spectral responses through controlled studies that systematically vary N availability across cultivars while monitoring chlorophyll, anthocyanin, and protein dynamics, thereby explaining the observed spectral patterns and enabling physics-informed machine learning approaches.

\section*{Acknowledgments}
This research was supported by the USDA through the National Institute of Food and Agriculture's (NIFA) Specialty Crop project award number: 2020-51181-32159. Any opinions, findings, and conclusions expressed in this publication are those of the authors and do not reflect any view from USDA or WSU. We thank Patrick A Scharf, Alan Kawakami, Syed Usama Bin Sabir, and the staff from OSU for their skilled technical assistance. Atif Bilal Asad would also like to thank Higher Education Commission(HEC) Pakistan for sponsoring his study at Washington State University.

\bibliographystyle{elsarticle-harv} 
\bibliography{refs}
\end{document}